\documentclass{tlp}
\usepackage{epic}
\usepackage{eepic}
\usepackage{latexsym}
\usepackage{afterpage}

\newcommand{\nt}[1]{\textsf{#1}}
\newcommand{\func}[1]{\textsc{#1}}

\newcommand{\ourtop}{\top}
\newcommand{\ipred}{\$\texttt{ipred}\$}
\newcommand{\gpred}{\$\texttt{gpred}\$}
\newcommand{\svars}[1]{\$\texttt{old}(\mbox{$#1$})\$}
\newcommand{\sgrounds}[1]{\$\texttt{ground}(\mbox{$#1$})\$}

\newcommand{\short}[1]{}

\newcommand{\oo}{\mbox{$[\![$}}
\newcommand{\cc}{\mbox{$]\!]$}}

\newcommand{\ttnew}{\texttt{new}}
\newcommand{\ttold}{\texttt{old}}
\newcommand{\ttground}{\texttt{ground}}
\newcommand{\ntnew}{\nt{new}}
\newcommand{\ntold}{\nt{old}}
\newcommand{\ntground}{\nt{ground}}

\newenvironment{grammar}{%
         \begin{center} \small%
         $\begin{array}{rcll}
         }{%
         \end{array}$\end{center}\ignorespaces%
         }

\newenvironment{definition}{%
         \begin{defn} \rm%
         }{%
         $\Box$ \end{defn}\ignorespaces%
         }
\newenvironment{example}{%
         \begin{ex} \rm%
         }{%
         $\Box$\end{ex}\ignorespaces%
         }
\newtheorem{theorem}{Theorem}

\newtheorem{ex}[theorem]{Example}
\newtheorem{defn}[theorem]{Definition}

\newenvironment{ttline}{\begin{trivlist}\small
                         \tt\item}{\end{trivlist}}
\newenvironment{ttprog}{\begin{trivlist}\small
                         \tt\item\begin{tabbing}}%
                       {\end{tabbing}\end{trivlist}}

\newcommand{\Inst}{\emph{Inst}}
\newcommand{\fresh}{\emph{fresh}}
\submitted{6 December 2002}
\revised{16 March 2004}
\accepted{15 September 2004}

\begin{document}

\title[Checking Modes of HAL Programs]{Checking Modes of HAL Programs\footnote{A preliminary version of this
paper appeared under the title ``Mode Checking in HAL,''
        in the {\em Conference on Computational Logic (CL'2000)},
        London, June 2000.
}}

\author[M.\ Garc\'{\i}a de la Banda et al.]{
MARIA GARCIA DE LA BANDA, 
WARWICK HARVEY, 
KIM MARRIOTT \\
  School of Computer Science \& Software Engineering,
  Monash University, Australia \\
\email{\{mbanda,marriott\}@csse.monash.edu.au}
\email{wh@icparc.ic.ac.uk}
\and PETER J. STUCKEY \\
 Department of Computer Science \& Software Engineering,
  University of Melbourne, Australia\\
\email{pjs@cs.mu.oz.au}
\and
BART DEMOEN \\
Department of Computer Science, 
Catholic University Leuven, Belgium\\
\email{bmd@cs.kuleuven.ac.be}
}

\maketitle

\textbf{Note:} This article is to published in
\emph{Theory and Practice of Logic Programming}.
\copyright Cambridge University Press.

\begin{abstract}
Recent constraint logic programming (CLP) languages, such as HAL and Mercury,
require type, mode and determinism
declarations for predicates. This information allows the generation of
efficient target code and the detection of many errors at compile-time.
Unfortunately, mode checking in such languages is difficult. One of the main 
reasons is that, for each predicate mode declaration,
the compiler is required to appropriately re-order literals in the
predicate's definition.
The task is further complicated by the need to 
handle complex instantiations
(which interact with type declarations and 
higher-order predicates) and automatic initialization of solver variables.
Here we define mode checking
for strongly typed CLP languages
 which require reordering of clause body literals. 
In addition, we show how to handle a simple case of 
polymorphic modes by using the
corresponding polymorphic types. 
\end{abstract}

\section{Introduction}

While traditional logic and constraint logic programming (CLP) languages are
untyped and unmoded, recent languages such as Mercury~\cite{mercury} and
HAL~\cite{cp99,flops2002} require type, mode and determinism declarations for
(exported) predicates. This information allows the generation
of efficient target code (e.g.\ mode information can provide 
an order of magnitude
speed improvement~\cite{iclp99}), improves robustness and facilitates
efficient integration with foreign language procedures.  
Here we describe our experience with mode checking in the
HAL compiler. 

HAL is a CLP language designed
to facilitate ``plug-and-play'' experimentation with different solvers.
To achieve this it provides support for 
user-defined constraint solvers, global variables
and dynamic scheduling. 
Mode checking in HAL
is one of the most complex stages in the compilation.
Since predicates can be given multiple mode declarations,
mode checking is performed for each of these modes and 
the compiler creates a 
specialized \emph{procedure} for each mode (i.e.\
it performs multi-variant specialization).
Mode checking involves traversing each predicate mode declaration
to check that if the predicate is called
with the input instantiation specified by the mode declaration then the
following two properties are satisfied. First, the predicate mode
declaration is \emph{input-output correct}, that is, 
it is guaranteed that if the input instantiation satisfies this
declaration then the result is an output instantiation that
satisfies the declaration.
And second, the predicate is
\emph{call correct}, 
that is, if the input instantiation satisfies this
declaration then each literal occurring in the definition of the
predicate is called with an input instantiation satisfying
one of its declared modes.

Call correctness may require the compiler to 
re-order literals in the body of each rule, 
so that literals are indeed called with an appropriate input instantiation. 
Such reordering is essential in logic programming
languages which wish to support multi-moded predicates while,
at the same time, retaining a Prolog programming style in which 
a single predicate definition is provided for all modes of usage. 
And an important function of this reordering is to  
appropriately order the equalities inserted by the compiler during program 
normalisation for matching/constructing 
non-variable predicate arguments.
The need to reorder rule bodies is one reason why
mode checking is a rather complex task.
However, it is not the only reason. Three other issues 
exacerbate the difficulty of mode checking.
First, 
instantiations (which describe the possible states of program variables) 
may be very complex and
interact with the type declarations. 
Second, accurate mode checking of higher-order predicates
 is difficult.
Third, the compiler needs to handle automatic initialization 
of solver variables.

Although
mode inference and checking of logic programs has been a fertile research
field for many years, 
almost all research has focused on
mode checking/inference in traditional (and thus untyped) 
logic programming languages where
the analysis assumes the given literal ordering is fixed
and cannot assume that a program is type correct.
Thus, a main contribution of this paper is
a complete definition of mode checking
in the context of CLP languages which are strongly typed
and which may require reordering of rule body literals during mode
checking.

A second contribution of the paper is
to describe the algorithms for mode checking currently employed in the
HAL compiler. Since HAL and the logic programming language Mercury
share similar type and mode systems,\footnote{In part, because HAL is compiled 
to Mercury.} much of our description and formalization also applies to 
mode checking in Mercury (which has not been previously described\footnote{
Recently a thesis has been completed on Mercury mode checking~\cite{dmo}}). 
However, there
are significant differences between mode checking in the two languages.
In HAL  there is the need to
 handle automatic initialization of solver variables,
and, in general, complex modes (other than \texttt{in} and
\texttt{out}) are used more frequently since constraint solver variables are 
usually not ground.
Furthermore, determining the best reordering in HAL is more complex 
than in Mercury because the order in which constraints are 
solved can have a more
significant impact on efficiency \cite{3r}.
Also,  HAL handles a limited form of polymorphic mode checking.
On the other hand, Mercury's mode system allows the specification of
additional information about data structure liveness and usage.

The rest of the paper is organized as follows. In the following
section we review related work. Section \ref{sec:lang} provides an
informal view of the role of types, modes, and instantiations in the
HAL language. Its aim is to give insight into the more rigorous
formalization provided in section \ref{sec:trees} which introduces
type-instantiation grammars for combining type and instantiation
information as the basis for mode checking in HAL. Section
\ref{sec:basic} describes the basic steps performed for mode checking
HAL programs. Section \ref{sec:init} focuses on the automatic
initialization needed by the modes of usage of some predicates.
Section \ref{sec:higher} discusses mode checking of higher order
predicates and objects, while Section \ref{sec:poly} shows how to
handle simple polymorphic modes. Finally, Section \ref{sec:concl}
provides our conclusions and discusses some future work.

\section{Related Work}

Starting with~\cite{Mellish87,deb-acm89} there has been considerable
research into mode checking and inference in traditional logic
programming languages.  However, as indicated above, there are two fundamental
differences between that work and ours.

First, almost all research
assumes that mode analysis is not required to reorder clause bodies.
Second, while almost all
research has focused on untyped logic programming languages, mode
checking of HAL relies on predicates and program variables
having a single (parametric polymorphic) 
Hindley-Milner type and the type correctness of the program
with respect to this type.
Access to  type information allows us to handle
more complex instantiations than are usually considered in mode analysis and
also to handle  mode checking of higher-order
predicates in a more rigorous fashion: in most previous work 
higher-order predicates are largely ignored.

Another important difference is that we are dealing with constraint
logic programming languages in which program variables
need to be appropriately initialized before being sent to
some constraint solver as part of a constraint. 
Requiring explicit initialization of solver variables puts additional
burden on the programmer and makes it impossible to write
multi-moded predicate definitions for which different modes require 
different variable initialisations. We have consequently chosen for 
the HAL compiler to
{\em automatically} initialize solver variables, i.e.\ the compiler
generates initialization code whenever necessary.
In order to perform such automatic initialization
mode checking in HAL must track
which program variables are currently uninitialized (in our terminology are
\ntnew{}). 
Tracking of uninitialized variables 
also supports powerful optimizations which can greatly improve performance.
For this reason the Mercury mode checker also tracks uninitialized variables.

This need to track uninitialized program variables 
is a significant difference between mode checking
in the Mercury and HAL languages, and most logic
programming work on modes.  
It is not the same as tracking so-called ``free'' variables in traditional
logic programming: first free variables may be aliased to other variables,
something that is not possible with uninitialized variables, second,
uninitialized variables have to be tracked exactly:
%
%
the compiler must not fail to initialize a variable, neither should it
initialize a variable more than once.
We will now review  selected related work in detail.

The original work on mode checking in strongly typed logic languages
with reorderable clause bodies is that of~\cite{Zoltan87}, which gives
an informal presentation of a mode system based on types. This is
perhaps the closest work in spirit since it was the basis of mode
checking in Mercury. However, its mode system is much simpler than
ours and it does not consider higher-order predicates or the problems of
automatic initialization.  The remaining work does not consider compile-time
reordering.

Perhaps the most closely related work in traditional logic programming
language analysis is the early work of~\cite{gerda} which uses regular
trees to define types and instantiations, and uses 
these trees to perform mode inference. Main differences 
are that~\cite{gerda} does not consider
reordering or tracking uninitialized variables. Other 
more technical differences are that,
although we use deterministic tree grammars to formalize types, our
type analysis~\cite{typec} is based on a Hindley-Milner approach. A
key difference with this and other work such as that of~\cite{mal}
is that we describe instantiations for polymorphic types, including
higher-order objects.  Also, in~\cite{gerda}
depth restrictions are imposed to make the generated regular trees
finite. This is not needed in our approach. Finally, they use definite
and possible sharing analysis to improve instantiation information.
This is not done yet in HAL for complexity reasons (sharing analysis
is quite expensive and thus a danger for practical compilation), however
a simple sharing and aliasing analysis should indeed prove to be
useful.

After the early work of~\cite{gerda}, there has been a significant
amount of research aimed at improving the precision of the analysis by
providing additional information about the structure of the terms.
Initially, this was achieved by performing some simple pattern
analysis and then providing this information to other analyses (see
for example, \cite{LeCharlier94:toplas,mulkers-iclp95}). Later, with
the gradual success of typed languages, pattern information was
substituted by type information with which more accurate results could
be obtained, i.e., type information was annotated with different kinds
of information some of which were mode information (see, for example,
\cite{rid-boiz99,smaus}). But most of this work was designed to either
provide a general framework for combining type information with other
kinds of information, or to infer some particular kind of information
(such as mode information) from a program without reordering the
literals in the body of predicates. Furthermore, they were not
interested in tracking uninitialized variables  nor keeping
enough instantiation information 
(i.e.\ which particular tree constructors can occur)
for optimizations such as switch
detection~\cite{mercury-switch}. Again, further differences arise
since we consider higher-order mode inference and polymorphic modes.

Recent work on \emph{directional types}~(see e.g.~\cite{mal}) 
is much more analogous to HAL mode checking.  
There, they are interested in determining mode-correctness of a program
given (user supplied) mode descriptions (called directional types).
Apart from previously mentioned differences, the framework
of~\cite{mal} uses directional types that are much simpler than
the instantiations that we deal with here.
Interestingly, 
the work of~\cite{mal} uses directional type correctness to show that
a run-time reordering of a well-typed program will not deadlock,
somewhat analogous to our compile-time reordering.

Type dependency analysis~\cite{mike} is also related to mode checking.
Their analysis determines 
type dependencies from which we can read all the 
correct modes or directional types of a program.  
The framework is however restricted to use types (and modes) defined by 
unary function symbols and an ACI operator.

Other related work has been on mode checking for concurrent logic
programming languages \cite{Codognet90}: 
There the emphasis has been on detecting
communication patterns and possible deadlocks.

The only other logic programming system we are aware of
which does significant mode checking is Ciao~\cite{CIAO}.
The Ciao logic programming system~\cite{CIAO} does mode
checking using its general assertion checking framework
CiaoPP based on abstract interpretation~\cite{SAS-CIAO}. 
Modes are considered as simply one form of assertion,
and indeed the notion of what is a mode is
completely redefinable.
The default modes are 
analyzed by the CiaoPP preprocessor using a combination of
regular type inference and groundness, freeness
and sharing analyses.  Ciao modes are more akin to directional types,
than the strong modes of HAL and Mercury, and the compiler will check
them if possible, and optionally add run-time tests for modes that
could not be checked at compile time.  
As with other earlier
work the fundamental differences with the HAL mode system are
in treatment of uninitialized variables, reordering, higher-order and
polymorphic modes.

\section{HAL by example}\label{sec:lang}

This section provides an informal view of the role of types, modes, and
instantiations in the HAL language. The aim is to provide insight into
the more rigorous formalization that will be provided in the following
sections. We do this by explaining the 
example HAL program shown in Figure~\ref{fig:stack}, which implements a
polymorphic stack using lists. Note that HAL follows the basic CLP syntax,
with variables, rules and predicates defined as usual (see, for example,
\cite{CLPbook} for an introduction to CLP).

\begin{figure}

\begin{center}
\begin{ttprog}
:- typedef list(T) -> ([] ; [T|list(T)]). \\
\\
:- instdef elist -> []. \\
:- instdef list(I) -> ([]; [I|list(I)]). \\
:- instdef nelist(I) -> [I|list(I)]. \\
\\
:- modedef out(I) -> (new -> I). \\
:- modedef in(I) -> (I -> I). \\
\\
:- pred push(list(T),T,list(T)). \\
:- mode push(in,in,out(nelist(ground))) is det. \\
push(S0,E,S1) :- S1 = [E|S0]. \\
\\
:- pred pop(list(T),T,list(T)). \\
:- mode pop(in,out,out) is semidet. \\
:- mode pop(in(nelist(ground)),out,out) is det. \\
pop(S0,E,S1) :- S0 = [E|S1]. \\
\\
:- pred empty(list(T)). \\
:- mode empty(in) is semidet. \\
:- mode empty(out(elist)) is det. \\
empty(S) :- S = [].  \\
\end{ttprog}
\end{center}
\caption{Example HAL program implementing a polymorphic stack.\label{fig:stack}}

\end{figure}

\subsection{Types}

%
%
Informally, a ground type describes a set of ground
terms and is used as a reasonable approximation of the
ground values a particular program variable can take. It is therefore
an invariant over the life time of the variable.
Types in HAL are prescriptive rather than descriptive, they restrict
the possible values of a variable.
Unlike much of the work performed on types for logic programming languages, 
our types only include the
ground (also called fixed) values that a variable can take.  
Later we will describe how instantiations are used to express when a 
variable takes a value which is not completely fixed.

Types are specified using type definition statements. 
For instance, in the example shown in
Figure~\ref{fig:stack}, the line
\begin{ttline}
:- typedef list(T) -> ([] ; [T|list(T)]).
\end{ttline}
defines the polymorphic type constructor \texttt{list/1} where \texttt{list(T)} is the
type of lists with elements of parametric type\footnote{In order to clearly distinguish
 between program variables and any other kinds of variables (type variables, instantiation variables, etc) we will refer to all other kinds of variables as parameters (i.e.,
type parameters, instantiation parameters, etc).} \texttt{T}. These lists are
made up using the \texttt{[]/0} and \texttt{./2} (represented by
\texttt{[$\cdot$|$\cdot$]}) tree constructors. 

HAL includes the usual set of built-in basic types: \texttt{float} 
(floating point numbers), \texttt{int} (integers), \texttt{char}
(characters) and \texttt{string} (strings). Like most typed languages, HAL
provides the means to define type equivalences. For example, the statement
\begin{ttline}
:- typedef vector = list(int).
\end{ttline}
defines the type \texttt{vector} to be a list of integers. Equivalence
types are simply macros for type expressions, and the compiler replaces
equivalence types by their definition (circular type equivalences are not
allowed). From now on we assume that equivalence types have been eliminated
from the type expressions we consider. This can be achieved
straightforwardly by applying substitution.

Finally, HAL allows a type to be declared 
as \emph{hidden} so that its definition
is not visible outside the module in which it is defined.  We note that the
treatment of hidden types is almost identical 
to that of type parameters and so omit
them for simplicity.

It is important to note that a program variable's type is used by 
a compiler to determine the
\emph{representation format} for that variable, i.e., the particular way
in which program variables are stored during execution. As a result,
%
%
two program variables may have different types
even though the representation of their values can be identical.
For example, in a language
providing both the ASCII character set and an extended
international character set, variables
representing each kind of character would need to have different types
since their internal representation is different. 

\subsection{Solvers}

In HAL a constraint solver is defined using a new type.
Assume for example, that a programmer wishes 
to implement a constraint solver over floating point numbers. 
From the point of view of the user, the variables will take
floating point values and 
thus one might expect them to have the built-in type
\texttt{float}.  
But their internal representation cannot be a float
as they need to keep track of internal information
for the solver.
As a result, the type of the variables cannot
be the built-in type \texttt{float} but must be some other type defined by the
solver, and whose implementation is hidden from the outside world.
This is were we use abstract types, to hide this view from the
outside world. 

\begin{example}\label{ex:cfloat}
For example a floating point solver type \texttt{cfloat}
might be defined as
\begin{ttline}
:- typedef cfloat -> var(int) ; val(float). 
\end{ttline}
where the integer in the \texttt{var} tree constructor 
refers to a column number in a (global) simplex tableaux,
and the \texttt{val} constructor is used to represent simple
fixed value floating point numbers.
\end{example}

Types defined by solvers 
are called \emph{solver types} and variables with a solver type are called
\emph{solver variables}.
Solvers must also provide an initialization procedure
(\texttt{init/1}) and at least the equality (\texttt{=/2}) constraint for
the type, although many other constraints will be usually
provided. Note that solver variables must be initialized before they can be
involved in any constraint. This is required so that the solver can
keep track of its variables and initialize the appropriate internal data-structures
for them.

The case of Herbrand solver types (i.e., types for which there is a
full unification solver) is somewhat special. Any user-defined type
can be declared to be a Herbrand solver type by annotating its type
definition with the words ``deriving solver''. For
example:
\begin{ttline}
:- typedef hlist(T) -> ([] ; [T|hlist(T)]) deriving solver.
\end{ttline}
defines the \texttt{hlist} Herbrand type. The compiler will then
automatically create an initialization predicate for the type (which
is actually identical for all Herbrand types) and an equality
predicate for the type which handles not only simple construction,
deconstruction and assignment of non-variable terms (which is the only
equality support provided for non-solver types), 
but full unification. As a
result, while variables with non-solver type \texttt{list} are always
required to be bound at run-time to a list of fixed length (so that the
limited support provided by construction, deconstruction and
assignment is enough),\footnote{Note that the elements inside the list need
  not be ground!} 
variables with type \texttt{hlist} may be bound to open
ended lists, where the tail of the list is an unbound (list)
variable.

\subsection{Instantiations}

Instantiations define the set of values, 
within a type, that a program variable may have at a
particular program point in the execution, as well
as the possibility that the variable (as yet) takes no value.
Instantiation information is vital to the compiler to determine
whether equations on terms are being used to
construct terms, deconstruct terms or check the equality of
two terms. Furthermore, instantiation information is needed to infer 
the determinism of predicates (i.e., how many answers a predicate has) 
and to perform many other low-level optimizations.

Although instantiations may seem very similar to types, 
they should not be confused: a type
is invariant over the life of the variable, while instantiations 
change.  Additionally, instantiations reflect the possibility
of a variable having no value yet, 
or being ``constrained'' to some unknown set of values.

HAL provides three \emph{base} instantiations for
a variable: {\tt ground}, {\tt old} and {\tt new}. A variable is
\texttt{ground} if it is known to have a unique value; the compiler might
not know exactly which value within the type (it might depend on the
particular execution), but it knows it is fixed (for a solver variable this
happens whenever the variable cannot be constrained further). 

A variable is \texttt{new} if it has not been initialized and it has
never appeared in a constraint (thus the name \texttt{new}). Thus, it
is known to take no value yet.  
As we have indicated, the instantiation \texttt{new} leads
to a crucial difference between mode checking in Mercury and HAL, 
and that investigated in most other research into mode checking
of logic programs.  Mercury and HAL demand that at each point
in execution the compiler knows whether a variable has a value or not.
This allows many compiler optimizations, and is a key to the
difference in execution speed of Mercury and HAL to most other logic
programming systems. The requirement to always have accurate
instantiation information about which variables are \texttt{new}
drives many of the decisions made in the mode checking system.
In particular, it means that a \texttt{new} variable is not allowed
to appear inside a data structure, and can only be given a value by
assignment or, if it is a solver variable, after initialization.

Finally, the instantiation \texttt{old} is used to describe a solver
variable that has been initialized but for which nothing is known
about its possible values.  Note that the variable might be
unconstrained, it might be ground, or anything in between (e.g.~be
greater than 5); the compiler simply does not know.  In the case in
which \texttt{old} is associated with a non-solver variable, it is
deemed to be equivalent to \texttt{ground}.  Note that in Mercury,
where there are no solver types, each variable always has an
instantiation which is either \texttt{new} or (a subset of)
\texttt{ground}.

It is important to note that \texttt{new} is not analogous to free in
the usual logic programming sense. A free variable in the HAL context
is an \texttt{old} variable (thus, it has been initialized by the
appropriate solver) which has never been bound to a non-variable term.
Thus, free variables might have been aliased, while \texttt{new}
variables cannot.
%
%
This is exploited by the compiler by not giving a run-time
representation to new variables. As a consequence, a \texttt{new}
variable cannot occur syntactically more than once.
%

For data structures such as trees or lists of solver variables, more
complex instantiation states may be used. These instantiations are
specified using instantiation definition statements 
which look very much like type definitions, the only
difference being that the arguments themselves are instantiations rather
than types.
For instance, in
the example shown in Figure~\ref{fig:stack}, the lines
\begin{ttline}
:- instdef elist -> []. \\
:- instdef list(I) -> ([] ; [I|list(I)]). \\
:- instdef nelist(I) -> [I|list(I)]. 
\end{ttline}
define the instantiation constructors \texttt{elist/0}, \texttt{list/1}
and \texttt{nelist/1}, which in the example are associated with variables of
type \texttt{list/1}. In that context, the instantiation \texttt{elist} describes
empty lists. The polymorphic instantiation \texttt{list(I)} describes lists
with elements of parametric instantiation \texttt{I} (note the deliberate reuse of the
type name). Finally, the instantiation \texttt{nelist(I)} describes
non-empty lists with elements of parametric instantiation \texttt{I}.

When associated with a variable, 
an instantiation requires the variable to be
bound to one of the outer-most 
functors in the right-hand-side of its
definition, and the arguments of the functor 
to satisfy the instantiation of the
corresponding arguments in the instantiation definition. 
In the case of \texttt{elist}, it would
mean the variable is \texttt{ground}. In the remaining two cases, it would
depend on the parametric instantiation \texttt{I}, but at the very least the variable
would be known to be a nil-terminated list, i.e.\ its length is fixed.

Note that the separation of instantiation information from type information
means we can associate the same instantiation for different types.
For example, a program variable with solver type \texttt{hlist(int)} and instantiation 
\texttt{list(ground)} indicates that the program variable 
has a fixed length list as its value.
A program variable with non-solver type \texttt{list(int)} and instantiation 
\texttt{list(ground)} indicates the same, but since the type is not a solver type, 
this would always be the case.
The separation of instantiation information from type information
also makes the handling of 
polymorphic application much more  straightforward,
since we will simply associate a different type with the same instantiation.

As mentioned before, the instantiation \texttt{new} is not allowed to appear as an
argument of any other instantiation. As a result, a variable can only be inserted
in a data structure if it is either \texttt{ground} or initialized (and
thus must be \texttt{old}). The main reason for this is the requirement
for accurate mode information about \texttt{new} variables. It quickly
becomes very difficult to \emph{always} have correct instantiation
information about which variables (and parts of data structures) are
\texttt{new}. 
While sharing and aliasing analyses might allow us
to keep track which variables are \texttt{new} in more situations,
inevitably they lead to situations where we cannot determine whether
the value of a variable is \texttt{new} or not, which is not acceptable to
the compiler.  We do however plan to use sharing and aliasing analysis
to keep track of initialized (\texttt{old}) variables that 
have yet to be constrained (analogous to free variables in Prolog).

\subsection{Modes}

A mode is of the form $\Inst_1 \rightarrow \Inst_2$ where $\Inst_1$ describes
the \emph{call} (or input) instantiation and $\Inst_2$ describes the
\emph{success} (or output) instantiation. 
The {\em base} modes are mappings from one base instantiation to another:
we use two letter codes (\texttt{oo}, \texttt{no}, \texttt{og},
\texttt{gg}, \texttt{ng}) 
based on the first letter of the instantiation, e.g.~\texttt{ng} 
is \texttt{new}$\rightarrow$\texttt{ground}.
The usual modes \texttt{in}  and
\texttt{out} are also provided (as renamings of \texttt{gg} and \texttt{ng}, respectively).

Modes are specified using mode definition statements. For instance, in the
example shown in Figure~\ref{fig:stack}, the lines
\begin{ttline}
:- modedef out(I) -> (new -> I). \\
:- modedef in(I) -> (I -> I).
\end{ttline}
are mode definitions, defining macros for modes. 
The \texttt{out(I)}
mode requires a new object on call and returns an object with instantiation
\texttt{I}. The \texttt{in(I)} mode requires instantiation \texttt{I} on
call and has the same instantiation on success.

HAL allows the programmer to define mode equivalences and instantiation
equivalences. As for type equivalences, from now on we assume that these
equivalences have been eliminated from the program. 
For example
\begin{ttline}
:- modedef in = in(ground). \\
:- modedef out = out(ground).
\end{ttline}
define \texttt{in} as equivalent to \texttt{ground -> ground},
and \texttt{out} as \texttt{new -> ground}.

\subsection{Equality}

The equality constraint is a special predicate in HAL.  
Equality will be normalized in HAL programs to take one of two
forms $x_1 = x_2$, and $x = f(x_1, \ldots, x_n)$ where
$x, x_1, \ldots, x_n$ are variables. 
Each form of equality supports a number of modes.

The equality $x_1 = x_2$ can be used in two modes.
In the first mode, \textbf{copy} (\texttt{:=}), either $x_1$
or $x_2$ must be \textsf{new} and the other variable must not be
\textsf{new}.  Assuming $x_1$ is new the value of $x_2$ is copied
into $x_1$. In the second mode \textbf{unify} (\texttt{==}) 
both $x_1$ and $x_2$ must not be new. This requires a full unification.

The equality $x = f(x_1, \ldots, x_n)$ can also be used in
two modes.  In the first mode, \textbf{construct} (\texttt{:=}),
$x$ must be \textsf{new} and each of $x_1, \ldots, x_n$ not \textsf{new}.
A new $f$ structure is built on the heap, and the values
of $x_1, \ldots, x_n$ are copied into this structure.
In the second mode, \textbf{deconstruct} (\texttt{=:}),
each of $x_1, \ldots, x_n$ must be \textsf{new} and $x$ must not be
\textsf{new}. If $x$ is of the form $f(a_1, \ldots, a_n)$ then
the value of $a_i$ is copied into $x_i$,
otherwise the deconstruct fails.\footnote{This is a
simplistic high level view, 
actually the system uses PARMA bindings and things are more complicated.
See~\cite{iclp99} for details.}
As we shall see later, mode checking shall extend the use
of these modes for other implicit modes.

\subsection{Predicate Declarations}

HAL allows the programmer to declare the type and modes
of usage of predicates. In our example of Figure~\ref{fig:stack}, the lines
\begin{ttline}
:- pred pop(list(T),T,list(T)). \\
:- mode pop(in,out,out) is semidet. \\
:- mode pop(in(nelist(ground)),out,out) is det. 
\end{ttline}
give such declarations for predicate \texttt{pop/3}.  
The first line is a
polymorphic type declaration (with parametric type \texttt{T}). It specifies the
types of each of the three arguments of \texttt{pop/3}.  The second and
third lines are mode declarations specifying the two different modes in
which the predicate can be executed. For example, in the 
first mode the first argument
is \texttt{ground} on call and success, while the second and third
arguments are \texttt{new} on call and \texttt{ground} on success.

Each mode declaration for a predicate defines a \emph{procedure},
a different way of executing the predicate. The role of mode checking
is not just to show these modes are correct, but also to reorder
conjunctions occurring in the predicate definition in order to
create these procedures.

The second and third lines also contain a determinism declaration.
These describe how many answers a predicate may have for a particular mode of usage:
\texttt{nondet} means any number of solutions; \texttt{multi} at least one
solution; \texttt{semidet} at most one solution; \texttt{det} exactly one
solution; \texttt{failure} no solutions; and \texttt{erroneous} a run-time
error.
 Thus, in the second line, since \texttt{pop/3} for this
mode of usage is guaranteed to have at most one solution  but might fail 
(when the first argument is an empty list), the determinism
is \texttt{semidet}.
For the second mode, the first argument is not only known to be ground but
also to be a non-empty list. As a result, the predicate can be ensured to
have exactly one solution and so its determinism is \texttt{det}.
Notice how by providing more complex instantiations we can improve the 
determinism information of the predicate.  They also lead to
more efficient code, since unnecessary checks (e.g.~that the
first argument of \texttt{pop/3} is bound to \texttt{./2}) 
are eliminated.


Currently, HAL requires predicate mode declarations for each predicate and
checks they are correct. Predicate type declarations, on the other hand,
can be omitted and, if so, will be inferred by the compiler~\cite{typec}.

\section{Type, Instantiation and Type-Instantiation Grammars}\label{sec:trees}

In this section we formalize
type and instantiation definitions in terms of (extended) regular
tree grammars. 
Then we introduce type-instantiation (ti-) grammars
which combine type and instantiation information
and are the basis for mode checking in HAL.
Throughout the section we will use \texttt{teletype}
font when referring to (fixed) type and
instantiation expressions, and 
\textsf{sans serif} font when referring
to non-terminals of tree grammars.

\subsection{HAL Programs}

We begin by defining basic terminology and HAL programs.

A \emph{signature} 
$\Sigma$ is a set of pairs $f/n$ where $f$ is a \emph{function symbol} and $n
\geq 0$ is the integer \emph{arity} of $f$.  
A function symbol with 0 arity is called a \emph{constant}.
Given a signature $\Sigma$ the set of
all \emph{trees} (the Herbrand Universe), 
denoted $\tau(\Sigma)$, is defined as the least set satisfying:
$$
\tau(\Sigma) = \bigcup_{f/n \in \Sigma} \{ f(t_1, \ldots, t_n) ~|~ 
                \{t_1, \ldots, t_n\} \subseteq \tau(\Sigma) \}.
$$
We assume (for simplicity) that $\Sigma$ contains at least one constant
symbol (i.e.\ a symbol with arity 0).

Let $V$ be a set of symbols called \emph{variables}. 
The set of all \emph{terms} over $\Sigma$ and $V$, denoted
$\tau(\Sigma, V)$, is similarly defined as the least set satisfying:
$$
\tau(\Sigma, V) = V \cup \bigcup_{f/n \in \Sigma} \{ f(t_1, \ldots, t_n) ~|~ 
                \{t_1, \ldots, t_n\} \subseteq \tau(\Sigma, V) \}
$$

A \emph{substitution} over signature $\Sigma$ and
variable set $V$ is a mapping from variables to terms in $\tau(\Sigma,V)$,
written $\{x_1/t_1, \ldots, x_n/t_n\}$.
We extend substitutions to map terms in the usual way.
A \emph{unifier} for two terms $t$ and $t'$ is a substitution 
$\theta$ such that $\theta(t)$ and $\theta(t')$ are syntactically identical.
A \emph{most general unifier} of two terms $t$ and $t'$, denoted $mgu(t,t')$
is a unifier $\theta$ such for every other unifier $\theta'$ of $t$ and $t'$
there exists a substitution $\theta''$ such that $\theta'$ is the composition
of $\theta$ with $\theta''$. 
Note that the only substitutions we shall deal with are
over type and instantiation parameters.

As we will be dealing with programs, types and instantiations there
will be a number of signatures of interest.
Let $V_{prog}$ be the set of program variable symbols,
and $\Sigma_{tree}$ be the tree constructors appearing in the program,
and $\Sigma_{pred}$ be the predicate symbols appearing in the program.
Let $V_{type}$ and $\Sigma_{type}$ be the
type variables and type constructors, and similarly let
$V_{inst}$ and $\Sigma_{inst}$ be the instantiation variables
and instantiation constructors. Note that these alphabets may overlap.

An \emph{atom} is of the form $p(s_1, \ldots, s_n)$ 
where $\{s_1, \ldots, s_n\} \subseteq \tau(\Sigma_{tree}, V_{prog})$
and $p/n \in \Sigma_{pred}$.
A \emph{literal} is either an atom, 
a variable-variable equation $x_1 = x_2$ where
$\{x_1, x_2\} \subseteq V_{prog}$, or a variable-functor equation 
$x = f(x_1, \ldots, x_n)$ where $f/n \in \Sigma_{tree}$ and
$x, x_1, \ldots, x_n$ are distinct elements of $V_{prog}$.
A \emph{goal} $G$ is a literal, a conjunction of goals
$G_1, \ldots, G_n$, a disjunction of goals $G_1 ; \cdots ; G_n$
or an if-then-else 
$G_i \texttt{~->~} G_t ; G_e$ (where $G_i, G_e, G_t$ are goals). 
A \emph{predicate definition} is of the form
$A \texttt{~:-~} G$ where $A$ is an atom and $G$ is a goal.

Note that we are assuming the programs have been normalized,
so that each literal has distinct
variables as arguments, each equality is either of the form $x_1 = x_2$ 
or $x = f(x_1,\ldots,x_n)$, where $x,x_1,\ldots,x_n$ are distinct variables,
and multiple bodies for a single predicate have been replaced by 
one disjunctive body. 

A \emph{predicate type declaration} is of the form
\begin{center}\tt
:- pred $p(t_1, \ldots, t_n)$
\end{center}
where
$\{t_1, \ldots, t_n\} \subseteq \tau(\Sigma_{type},V_{type})$ are
type expressions.
A \emph{predicate mode declaration} is of the form
\begin{center}\tt
:- mode $p(c_1 \rightarrow s_1, \ldots, c_n \rightarrow s_n)$ 
\end{center}
where 
$\{c_1, \ldots, c_n, s_1, \ldots, s_n\} 
\subseteq \tau(\Sigma_{inst})$ are ground
instantiation expressions.
A \emph{complete predicate definition} for  predicate  symbol $p/n \in \Sigma_{pred}$
consists of a
predicate definition, a predicate type declaration, and a non-empty set of
predicate mode declarations for  $p/n$.
A \emph{program} is a collection of complete predicate definitions
for distinct predicate symbols.

\subsection{Tree Grammars}

Tree grammars 
are a well understood formalism (see, for example, \cite{gs,tata97})
for defining regular tree languages.
We first review the standard definitions for tree grammars
since we shall have to
extend these
in order to handle the complexities of mode checking.
 
A \emph{tree grammar} 
$r$ over signature $\Sigma$ and
\emph{non-terminal set} $NT$ 
is a finite set of \emph{production rules} of 
the form $x \rightarrow t$ where
$x \in NT$ and $t$ is of the form $f(x_1, \ldots, x_n)$
where $f/n \in \Sigma$ and $\{x_1, \ldots, x_n\} \subseteq NT$.
For each $x \in NT$ and $f/n \in \Sigma$ we require
that there is at most one rule of the form
$x \rightarrow f(x_1, \ldots, x_n)$; hence the grammars
are \emph{deterministic}.

We have chosen to restrict ourselves to deterministic tree grammars:
these grammars are expressive enough for Hindley-Milner types and they
give rise to simpler, more efficient algorithms---an important
consideration for a compiler designed for large real-world programs.

We assume that from a grammar $r$ 
we can determine its \emph{root non-terminal},
denoted $root(r)$.  In reality this is an additional piece of
information attached to each grammar.
We shall write grammars so that the root non-terminal
appears on the left hand side of
the first production rule in $r$.

It will often be useful to extract a sub-grammar $r'$ from a 
grammar $r$ defining some non-terminal $x$ appearing in $r$.  
If $x$ is a non-terminal occurring in grammar
$r$, then $subg(x,r)$ is the set of rules in $r$ for $x$ and
all other non-terminals reachable from $x$. Or more precisely, 
$subg(x,r)$ is the smallest 
set of rules satisfying
\begin{eqnarray*}
subg(x,r) & \supseteq & \{ x \rightarrow t \in r \} \\
subg(x,r) & \supseteq & \{ x' \rightarrow t \in r ~|~ x' \in NT, 
                      \exists x'' \rightarrow g(x''_1, \ldots, x', \ldots,
                   x''_m) \in subg(x,r)\}
\end{eqnarray*}
The root of the grammar $subg(x,r)$ is $x$, i.e.\ $root(subg(x,r)) = x$.

\begin{example}\label{ex:gs}
Consider the signature 
$\{[]/0, \textrm{`}.\textrm{'}/2, a/0, b/0, c/0, d/0\}$ and the
non-terminal set \{\nt{abc}, $\nt{list(abc)}$, $\nt{bcd}$, 
$\nt{evenlist(bcd)}$, $\nt{oddlist(bcd)}$\}, 
then two example regular tree
grammars over this signature and non-terminal set are $r_1$:
\begin{grammar}
\nt{list(abc)} & \rightarrow & [] \\
\nt{list(abc)} & \rightarrow & [\nt{abc} | \nt{list(abc)}] \\
\nt{abc} & \rightarrow & a \\
\nt{abc} & \rightarrow & b \\
\nt{abc} & \rightarrow & c \\
\end{grammar}
and $r_2$:
\begin{grammar}
\nt{evenlist(bcd)} & \rightarrow & [] \\
\nt{evenlist(bcd)} & \rightarrow & [ \nt{bcd} | \nt{oddlist(bcd)} ] \\
\nt{oddlist(bcd)} & \rightarrow & [ \nt{bcd} | \nt{evenlist(bcd)} ] \\
\nt{bcd} & \rightarrow & b \\
\nt{bcd} & \rightarrow & c \\
\nt{bcd} & \rightarrow & d \\
\end{grammar}
The root non-terminal of $r_1$ is \nt{list(abc)}, while
the root non-terminal of $r_2$ is \nt{evenlist(bcd)}.
The grammar $subg(\nt{abc}, r_1)$ consists of  the last three rules of $r_1$
while the grammar $subg(\nt{oddlist(bcd)}, r_2)$ includes all of the
rules of $r_2$ but we would write the third rule in the first position,
to indicate the root non-terminal was \nt{oddlist(bcd)}.
\end{example}

A production of form  $x \rightarrow s$ in some
grammar $r$ can be used to rewrite a term $t \in \tau(\Sigma,NT)$
containing an occurrence of $x$  to the term $t' \in \tau(\Sigma,NT)$ 
where $t'$ is obtained by
replacing the occurrence of $x$ in $t$ by $s$. This is called
a  \emph{derivation step} and is denoted by
$t \Rightarrow t'$. 
We let $\Rightarrow^*$ be the transitive,
reflexive closure of $\Rightarrow$.
The \emph{language generated} by $r$, denoted by $\oo{} r \cc{}$,
is the set 
$$\{ t \in \tau(\Sigma) \mid  root(r) \Rightarrow^* t \}$$

\begin{example}
For example, consider the grammars of Example~\ref{ex:gs}.
The set $\oo r_1 \cc$ is all lists of $a$'s, $b$'s and $c$'s, while
$\oo r_2 \cc$ is all even length lists of 
$b$'s, $c$'s and $d$'s.
\end{example}

For brevity we shall often write tree grammars in a more compressed form.
We use
$$
x \rightarrow t_1 ; t_2 ; \cdots ; t_n
$$
as shorthand for the set of production rules:
$x \rightarrow t_1$, $x \rightarrow t_2$, \ldots, $x \rightarrow t_n$.

The $\oo{} \cdot \cc{}$ function 
induces a pre-order on 
tree grammars: $r_1 \preceq r_2$ 
iff $\oo{} r_1 \cc{} \subseteq \oo{} r_2 \cc{}$. 
If
we regard grammars with the same language as equivalent,  
$\preceq$ gives rise
to a natural partial order over these equivalence classes
of tree grammars. In fact they form a lattice.
However, we shall largely ignore 
these equivalence classes since all of our operations work on concrete
grammars. 

We shall also make use of two special grammars. The first is the
\emph{least tree grammar}, which we denote by $\bot$.
We define that $\oo \bot \cc = \emptyset$, and so, as its name suggests
we have that $\bot \preceq r$ for all grammars $r$.
During mode checking the $\bot$ grammar indicates where 
execution is known to fail. 
The second special grammar is the \emph{error grammar},
denoted by $\ourtop$. It is used to indicate that a mode error has occurred
and we define that $r \preceq \ourtop$ for all tree grammars $r$.
 
We use $\sqcap$ to 
denote the meet (i.e.\ greatest  lower bound) operator 
on grammars, 
and $\sqcup$ to denote the join (i.e.\ least upper bound)
operator. 
We assume that the non-terminals appearing in the two grammars
to be operated on are renamed apart. 

We have that  $\oo{} r_1 \sqcap
r_2 \cc{} = \oo{} r_1 \cc{} \cap \oo{} r_2 \cc{}$.  Because we restrict
ourselves to deterministic tree grammars 
the join is inexact: That is to say, 
$\oo{} r_1 \sqcup r_2 \cc{} \supseteq \oo{} r_1 \cc{} \cup
\oo{} r_2 \cc{}$, and for some $r_1$ and $r_2$,
$\oo{} r_1 \sqcup r_2 \cc{} \ne \oo{} r_1 \cc{} \cup
\oo{} r_2 \cc{}$.
Of course, since it is the join, it is as precise as possible:
for any grammar $r$ such that 
$\oo{} r \cc{} \supseteq \oo{} r_1 \cc{} \cup\oo{} r_2 \cc{}$, we have that 
$\oo{} r \cc{} \supseteq \oo{} r_1 \sqcup r_2 \cc{}$.

Algorithms for determining if $r_1 \preceq
r_2$, and constructing $r_1 \sqcap r_2$ and $r_1 \sqcup r_2$
are straightforward and omitted.\footnote{The final operations of interest are
given in the appendix.}

\begin{example}
Consider the grammars  $r_1$ and $r_2$ of Example~\ref{ex:gs}.
Their meet $r_1 \sqcap r_2$ is:
\begin{grammar}
\nt{meet(list(abc),evenlist(bcd))} & \rightarrow & []  ~;~ [ \nt{meet(abc,bcd)} ~|~ \nt{meet(list(abc),oddlist(bcd))}] \\
\nt{meet(abc,bcd)} & \rightarrow & b   ~;~ c \\
\nt{meet(list(abc),oddlist(bcd))} & \rightarrow & [ \nt{meet(abc,bcd)} ~|~ \nt{meet(list(abc),evenlist(bcd))} ] \\
\end{grammar}
while their join $r_1 \sqcup r_2$ is:
\begin{grammar}
\nt{join(list(abc),evenlist(bcd))} & \rightarrow & []  ~;~ [ \nt{join(abc,bcd)} ~|~ \nt{join(list(abc),oddlist(bcd))}] \\
\nt{join(abc,bcd)} & \rightarrow & a ~;~ b  ~;~ c ~;~ d \\
\nt{join(list(abc),oddlist(bcd))} & \rightarrow & []  ~;~ [ \nt{join(abc,bcd)} ~|~ \nt{join(list(abc),evenlist(bcd))} ] \\
\end{grammar}
Note that the language generated by the grammar $r_1 \sqcup r_2$ 
could be represented with fewer rules. In the compiler there is no 
effort to build minimal representations of grammars since 
non-minimal grammars do not seem to occur that often in practice.
\end{example}

\subsection{Types} 

Types in HAL are polymorphic Hindley-Milner types.
\emph{Type expressions} (or \emph{types}) are 
terms in the language $\tau(\Sigma_{type}, V_{type})$
where $\Sigma_{type}$ are \emph{type constructors} and
variables $V_{type}$ are \emph{type parameters}. 
Each type constructor $f/n \in \Sigma_{type}$ must have
a definition.

\begin{definition}
A \emph{type definition} for $f/n \in \Sigma_{type}$ 
is of the form

\begin{center}
{\tt :- typedef~} $f(v_1, \ldots, v_n)$ {\tt ~->~} $(f_1(t_1^1,
  \ldots, t^1_{m_1}) ; \cdots ; f_k(t^k_1, \ldots, t^k_{m_k}))$. 
\end{center}

\noindent
where $v_1, \ldots, v_n$ 
        are distinct type parameters, 
$\{f_1/m_1, \ldots, f_k/m_k\}   \subseteq \Sigma_{tree}$
  are distinct tree constructor/arity pairs, 
  and $t_1^1, \ldots, t^k_{m_k}$ are type
  expressions involving at most parameters $v_1, \ldots, v_n$.
The type definition for $f/n$ 
may optionally have \texttt{deriving solver}
appended. If so then types of the form $f(t_1, \ldots, t_n)$
are \emph{solver types}, otherwise they are \emph{non-solver types}. 
\end{definition}

Clearly, the type definition for $f$ can be viewed 
as simply a set of production rules over signature $\Sigma_{tree}$ 
and non-terminal set
$\tau(\Sigma_{type},V_{type})$.

We can associate with each (non-parameter) 
type expression the production 
rules that define the topmost symbol of the type.
Let $t$ be a type expression of the form $f(t_1, \ldots, t_n)$
and let $f/n$ have type definition

\begin{center}
{\tt :- typedef~} $f(v_1, \ldots, v_n)$ {\tt ~->~} $(f_1(t_1^1,
  \ldots, t^1_{m_1}) ; \cdots ; f_k(t^k_1, \ldots, t^k_{m_k}))$. 
\end{center}

\noindent
We define $rules(t)$ to be the production rules:
$$\theta(f(v_1, \ldots, v_n)) \rightarrow (f_1(\theta(t_1^1),
  \ldots, \theta(t^1_{m_1})) ; \cdots ; f_k(\theta(t^k_1), \ldots,
\theta(t^k_{m_k})))$$
where $\theta = \{v_1/t_1, \ldots, v_n/t_n\}$.
If $t \in V_{type}$ we define $rules(t)$ to be the empty set.

We can extend this notation to associate a tree grammar with
a type expression. 
Let $grammar(t)$ be the least set of production rules such that:
\begin{eqnarray*}
grammar(t) & \supseteq & rules(t) \\
grammar(t) & \supseteq & \bigcup \{ rules(t') ~|~ \exists x' \rightarrow
g(t'_1, \ldots, t', \ldots, t'_m) \in grammar(t)) \}
\end{eqnarray*}
We assume that $root(grammar(t)) = t$.
Note at this point we make no distinction between solver types
and non-solver types; this will only occur once we consider
instantiations.

In order to avoid type expressions that
depend on an infinite number of types 
we restrict the type definitions to 
be \emph{regular}~\cite{mycroft}.
A type $t$ is \emph{regular} if $grammar(t)$ is finite.\footnote{Note that 
non-regular types are rarely used (although see~\cite{okasaki}). 
The compiler could be extended
to support mode checking 
for non-regular types as long as we keep the restriction to
regular instantiations.}

Consider for example the non-regular type definition:
\begin{ttline}
:- typedef erk(T) -> node(erk(list(T)), T).
\end{ttline}
The meaning of the 
type \texttt{erk(int)} depends on the meaning of the
type \texttt{erk(list(int))}, which depends on the meaning of the
type \texttt{erk(list(list(int)))}, etc. 
By restricting to regular types we are guaranteed that each type
expression only involves a finite number of types.

A \emph{ground type expression} $t$ 
is an element of $\tau(\Sigma_{type})$.
The grammar corresponding to ground type expression $t$ 
defines the meaning of the type expression as
a set of trees ($\oo grammar(t) \cc$). 
Note that during run-time every variable (for each invocation of a
predicate) has a unique ground 
type in $\tau(\Sigma_{type})$.

\begin{example}
Given the type definitions:
\begin{ttprog}
:- typedef abc -> a ; b ; c. \\
:- typedef list(T) -> [] ; [T | list(T)].
\end{ttprog}
then the grammar $r_1$ shown in Example~\ref{ex:gs} is 
$grammar(\texttt{list(abc)})$. 
The set $\oo r_1 \cc$ is the set 
of lists of \texttt{a}'s, \texttt{b}'s and \texttt{c}'s.
The grammar $grammar(\texttt{list(T)})$
is
\begin{grammar}
\nt{list($T$)} & \rightarrow & [] ~;~ [\nt{$T$}|\nt{list($T$)}] 
\end{grammar}
The set $\oo grammar(\texttt{list(T)}) \cc = \{ [] \}$.
\end{example}

Note that the grammars corresponding to non-ground type expressions are
not very interesting, as illustrated in the above example.
We can think of a non-ground type expression as a mapping from
grounding substitutions to (ground) types whose meaning is then given
by their corresponding grammar.

The built-in types \texttt{float}, \texttt{int}, \texttt{char} and
\texttt{string} are conceptually expressible as (possibly infinite) 
tree grammars. 
For example, \texttt{int} can be thought of as
having the (infinite) definition:
\begin{ttline}
:- typedef int -> 0 ; 1 ; -1 ; 2 ; -2 ; 3 ; ...
\end{ttline}
Though the infinite number of children will render some of the algorithms
on tree grammars ineffective this is easily avoided in the compiler 
by treating the type expressions specially (we omit details
in our algorithms since it is straightforward). 

Note that in HAL, type inference and checking is performed using a
constraint-based Hindley-Milner 
approach on the type expressions~\cite{typec}.  
In this paper we assume that type analysis
has been performed previously, and there are no type errors.
For the purposes of mode
checking the type correctness of a program has four main consequences.
First, each program variable is known to have a unique polymorphic
type. Second, all values taken 
by a variable during the execution are known to be
members of this type.  Third, calls to a polymorphic 
predicate are guaranteed to have an equal or more specific type than that
of the predicate. Fourth, all type parameters appearing in the type
of a variable in the body of a predicate are known to also appear in the type
of some variable in the head of the predicate. Together, these
guarantee that whenever we compare grammars during mode checking, they
correspond to exactly the same type.\footnote{Even mode checking
a call to a polymorphic 
predicate will use the calling type, which may be more specific than the
predicate's declared type.}  
This is used to substantially
simplify the algorithms for mode checking (see, for example, the re-definition of 
function $\preceq$ in Section~\ref{sec:type-inst}, and the assumption on the 
existence of type environment 
$\theta$ at the beginning of Section~\ref{sec:sched-ho}).

\subsection{Values}

Types only express sets of fixed values (subsets of $\tau(\Sigma_{tree})$).
However, during execution variables do not always have a fixed value
and it is the role of mode checking to track these changes in variable
instantiation.
Thus, in order to perform mode checking we need to introduce special 
constants, \#\texttt{fresh}\# and \#\texttt{var}\#, to represent the
two kinds of non-fixed values that a program variable can have during
execution.

The \#\texttt{fresh}\# constant is used to represent that a 
program variable takes no value (i.e.,~it has not been initialized),
and corresponds to the \ttnew{} instantiation.  
Note that in HAL there is no run-time representation for \#\texttt{fresh}\# 
variables. As a result, the compiler needs to know at all times 
whether a variable is \ttnew{} or not. Thus, 
any tree language including
\#\texttt{fresh}\# and some other term is not a valid
description of the values of a program variable.  

The \#\texttt{var}\# constant is used to represent a 
program variable (or part of a value) that has been initialized but
not further constrained. It corresponds to a ``free'' variable 
in the usual logic
programming sense. 
 The \#\texttt{var}\# constructor will arise in 
descriptions of \ttold{} instantiations, where we can define values which
are not fixed.  Of course it will only make sense for variables of 
solver types to take on this value. 

The values that a variable can take are thus represented by trees in
$\tau(\Sigma_{tree} \cup \{\#\texttt{var}\#\}) \cup \{ \#\texttt{fresh}\# \}$.

\subsection{Instantiations}

A type expression by itself represents a set of
fixed values. An instantiation by itself has little meaning, it is just a term
in the language of expressions. Its meaning is only defined when it
is considered in the context of a type expression.  For instance, the 
meaning of \ttground{} depends upon the type of the variable it is 
referring to.

In the following section we define a function $\func{rt}(t, i)$ 
which takes a type expression $t$ and an instantiation $i$ and 
returns a tree-grammar defining the set of (possibly non-fixed) 
values that a program variable with the given type and instantiation can
take. 
In this section we define 
the function $\func{base}(t, i)$ which is the function
$\func{rt}(t,i)$ for the particular case in which $i$ is a base instantiation.
In order to avoid name clashes, the
function creates a unique non-terminal grammar symbol
$ti(t,base)$ for the type $t$ and base instantiation $base$
with which it is called and returns this together with the
grammar for $t$ and $base$.
The symbol $ti(t,i)$ represents the root of the tree-grammar
which defines the possible values of a variable of type $t$ and
instantiation $i$.

When a program variable is \ttnew{} it can only have one possible
value, \#\texttt{fresh}\#. Hence the grammar returned by
$\func{base}(t, \ttnew)$ for any type $t$ is simply
\begin{grammar}
ti(t,\ntnew) & \rightarrow & \#\texttt{fresh}\#
\end{grammar}
In a slight abuse of notation we will use \ntnew{} to refer to this grammar.

When a program variable is \ttground{} it can take any fixed value.
If the type $t$ of the variable is ground, then $\func{base}(t,\ttground{})$ 
is identical to the grammar defining its type ($grammar(t)$). 
Type parameters complicate this somewhat. 
Since we are going to reason about
the values of variables with non-ground types we need a way of representing
the possible ground values of a type parameter. We introduce new 
constants of the form $\$\ttground(v)\$$ 
where $v \in V_{type}$ 
to represent these languages. 
So for $t \in V_{type}$ the grammar $\func{base}(t,\ttground)$ is defined as 
\begin{grammar} 
ti(t,\ttground) & \rightarrow & \$\ttground(t)\$
\end{grammar}
For arbitrary types $t$, $\func{base}(t,\ttground)$ is defined as
the union of the rules
\begin{grammar}
ti(t',\ttground)  & \rightarrow & f(ti(t_1,\ttground), \ldots,
 ti(t_n,\ttground))
\end{grammar}
for each $t' \rightarrow f(t_1, \ldots, t_n)$ occurring in $grammar(t)$,
with
\begin{grammar}
ti(t',\ttground) & \rightarrow & \$\ttground(t')\$ 
\end{grammar}
for each $t' \in V_{type}$ occurring in $grammar(t)$.

Conceptually, the new constant $\$\ttground(v)\$$
is a place holder for the grammar 
$\func{base}(t',\ttground)$ obtained if 
$v$ were replaced by the ground type $t'$.

When a program variable is \ttold{} it can take any initialized value.
This will have a different effect on the
parts of the type which are solver types themselves
and on those which are not.
Non-solver types do not allow the 
possibility of taking an initialized but unbound value 
(represented by the value \#\texttt{var}\#). 
Thus, for solver types $t$ we shall add a production rule 
$t \rightarrow \#\texttt{var}\#$ to the usual rules defining
the type, while non-solver types remain unchanged. 
In order to handle type parameters we introduce another set of
constants $\$\ttold(v)\$$ where $v \in V_{type}$. Each constant is simply a 
place holder for $\func{base}(t',\ttold)$ obtained if $v$ were replaced by 
the ground type $t'$. Thus, $\func{base}(t,\ttold)$ for $t \in V_{type}$ 
is defined as 
\begin{grammar} 
ti(\nt{t},\ttold) & \rightarrow & \$\ttground(t)\$ ~;~ \$\ttold(t)\$
\end{grammar}
and 
otherwise $\func{base}(t,\ttold)$ is defined as the rules
\begin{grammar}
ti(t',\ttold)  & \rightarrow & f(ti(t_1,\ttold), \ldots, ti(t_n,\ttold))
\end{grammar}
for each rule $t' \rightarrow f(t_1, \ldots, t_n)$ in $grammar(t)$,
together with
\begin{grammar}
ti(t',\ttold) & \rightarrow &\#\texttt{var}\#
\end{grammar}
for each solver type $t'$ occurring in $grammar(t)$,
and
\begin{grammar}
ti(t',\ttold) & \rightarrow & \$\ttground(t')\$ ~;~ \$\ttold(t')\$ 
\end{grammar}
for each type variable $t' \in V_{type}$ occurring in $grammar(t)$.

The reason we represent an \texttt{old} variable of type $t$ 
using both the $\$\ttground(t')\$$ and $\$\ttold(t')\$$,
is that then a \texttt{ground} variable of type $t$ defines
a sublanguage. This will simplify many algorithms.

\begin{example}\label{ex:labcs}
Given the type definitions:
\begin{ttprog}
:- typedef abc -> a ; b ; c. \\
:- typedef hlist(T) -> [] ; [T | hlist(T)] deriving solver.
\end{ttprog}
Then $\emph{olabc1} = \func{base}(\texttt{hlist(abc)},\texttt{old})$ is the 
grammar:
\begin{grammar}
ti(\nt{hlist(abc)},\ttold) & \rightarrow & []  ~;~~ [ ti(\nt{abc},\ttold) ~|~ ti(\nt{hlist(abc)},\ttold)] ~;~ \#\texttt{var}\# \\
ti(\nt{abc},\ttold) & \rightarrow & a ~;~ b   ~;~ c
\end{grammar}

The set $\oo{} \emph{olabc1} \cc{}$ 
includes the values $[], [a | \#\texttt{var}\# ],
[b], [b,a,c,a | \#\texttt{var}\# ]$.
The symbol \#\texttt{var}\# represents an uninstantiated 
variable, and so the second and fourth values are open-ended lists.

As another example,
imagine we swap which type is a solver type. That is,
suppose we have definitions
\begin{ttprog}
:- typedef habc -> a ; b ; c deriving solver. \\
:- typedef list(T) -> [] ; [T | list(T)].
\end{ttprog}
Then $\emph{olabc2} = \func{base}(\texttt{list(habc)},\texttt{old})$ is the grammar:
\begin{grammar}
ti(\nt{list(habc)},\ttold) & \rightarrow & []  ~;~~ [ ti(\nt{habc},\ttold)  ~|~ ti(\nt{list(habc)},\ttold)] \\
ti(\nt{habc},\ttold) & \rightarrow & a ~;~ b   ~;~ c ~;~ \#\texttt{var}\# 
\end{grammar}
The set $\oo{} \emph{olabc2} \cc{}$ 
includes the values $[], [a], [\#\texttt{var}\#, b, \#\texttt{var}\#]$ 
which are all fixed-length lists whose elements may be variables. 
Note that the two occurrences of the symbol \#\texttt{var}\#
in the last tree do not necessarily represent the same solver variable.

Finally $\func{base}(\texttt{hlist(T)}, \texttt{old})$ is (using the first
definition) 
\begin{grammar}
ti(\nt{hlist($T$)},\ttold) & \rightarrow & []  ~;~~ [ ti(T,\ttold) ~|~ ti(\nt{hlist($T$)},\ttold) ] ~;~ \#\texttt{var}\# \\
ti(T,\ttold) & \rightarrow &  \$\texttt{ground}(T)\$ ~;~ \$\ttold(T)\$
\end{grammar}
\end{example}

Let us now consider instantiations in general, rather than only base instantiations.
\emph{Instantiation expressions} (or \emph{instantiations}) 
are terms in the language $\tau(\Sigma_{inst}, V_{inst})$
where $\Sigma_{inst}$ are \emph{instantiation constructors} and
variables $V_{inst}$ are \emph{instantiation parameters}. 
Each instantiation 
constructor $g/n \in \Sigma_{inst}$ must have
a definition.
Often, we will overload functors as both type and 
instantiation constructors
(so $\Sigma_{type}$ and $\Sigma_{inst}$ are not disjoint).
The base instantiations (\texttt{ground}, \texttt{old} and
\texttt{new}) are simply special 0-ary elements of $\Sigma_{inst}$.

\begin{definition}
An \emph{instantiation definition} for $g$ is of the form:

\begin{center}
{\tt :- instdef~} $g(w_1, \ldots, w_n)$ {\tt ~->~}  $(g_1(i_1^1,
  \ldots, i^1_{m_1}) ; \cdots ; g_k(i^k_1, \ldots, i^k_{m_k})).$
\end{center}

\noindent
  where $w_1, \ldots, w_n$ are distinct instantiation 
  parameters, 
  $\{g_1/m_1, \ldots, g_k/m_k\} \subseteq \Sigma_{tree}$
  are distinct tree constructors, and $i_1^1, \ldots, i^k_{m_k}$ 
  are instantiation expressions other than 
\texttt{new}\footnote{As mentioned before, disallowing nesting
of the  \texttt{new} instantiation simplifies mode analysis. It also ensures that all subparts of 
a data structure have a proper representation at run-time.}
involving at most the parameters $w_1, \ldots, w_n$.
Just as for type definitions, we demand that 
instantiation definitions are \emph{regular}.\footnote{It is hard to 
see how to lift this restriction.}
\end{definition}

We can associate a set of production rules $rules(i)$ with an
instantiation expression $i$ just as we do for type expressions.
For the base instantiations we define $rules(\ttnew) = 
rules(\ttold) = rules(\ttground) = \emptyset$.

A \emph{ground instantiation} is an element of $\tau(\Sigma_{inst})$. 
The existence of instantiation parameters during mode analysis
would significantly complicate the task of the analyzer. 
This is mainly 
because functions to compare type-instantiations or to
compute their join and meet would need to
return a set of constraints involving instantiation parameters. 
Furthermore, predicate mode declarations containing
instantiation parameters might need to express some constraints involving
those instantiations. 
Therefore, for simplicity, HAL (like Mercury\footnote{Recently Mercury has
added a (as yet unreleased) 
feature allowing limited non-ground instantiations in
predicate modes.})  
requires instantiations
appearing in a predicate mode declaration to be ground. As a result,
mode checking only deals with ground instantiations and, from now on,
we will assume all instantiations are ground.

The reason this problem does not arise with type parameters is
that, as mentioned before, type correctness guarantees that 
whenever we compare type-instantiations, the two types being compared are syntactically 
identical. Thus, if two type parameters are being compared, they are guaranteed to
be the same type parameter.

\subsection{Type-Instantiation Grammars}\label{sec:type-inst}

In this section we define the 
function $\func{rt}(t,i)$ which takes a type expression $t$ and a
ground instantiation expression $i$ and returns a 
\emph{type-instantiation tree grammar} (or \emph{ti-grammar}).
Mode checking will manipulate \emph{ti-grammar}s, built
from the types and instantiations occurring in the program.

The function \func{rt} 
defines the meaning of combining a type with an instantiation
by extending \func{base} to non-base instantiations.
A non-base instantiation combines with a type in a manner
analogous to the $\sqcap$ operation over the rules defining
each other.  
Intuitively the function \func{rt} intersects the grammars
of $t$ and $i$.  This is not really the case because of special treatment
of type parameters and base instantiations.

Figure~\ref{fig:rt} gives the algorithm for computing $\func{rt}(t,i)$. 
The 
function $\func{rt}(t,i,P)$
 does all of the work.
It creates a unique grammar symbol
$ti(t,i)$ for the type $t$ and instantiation $i$
with which it is called and returns this with the
type instantiation grammar for $t$ and $i$.
Its last argument $P$ is the
set of grammar symbols constructed in the parent calls:
this is used to check that the symbol $ti(t,i)$
has not already been encountered and so avoids infinite recursion.
The root of the grammar $r$ returned is the symbol $ti(t,i)$.

Note that it is a mode error to associate a non-base instantiation
with a parameter type $t \in  V_{type}$, since we cannot know
what function symbols make up the type $t$. 
In this case the algorithm returns the special $\ourtop$ 
grammar indicating a mode error.

\begin{figure}

\begin{small}
\begin{tabbing}
xx \= xx \= xx \= xx \= xx \= xx \= xx \= xx \= \kill
\func{rt}($t$,$i$) \\
\> $(r,\_)$ := \func{rt}($t$,$i$,$\emptyset$) \\
\> \textbf{return} $r$ \\
\\
\func{rt}($t$,$i$,$P$) \\
\> \textbf{if} ($ti(t,i) \in P$) 
                \textbf{return} $(\emptyset,ti(t,i)))$\\
\> \textbf{if} ($i$ is a base instantiation) \textbf{return}
(\func{base}($t$,$i$), $ti(t,i)$) \\
\> \textbf{if} ($t \in V_{type}$) \textbf{return} $(\ourtop, \_)$ \\
\> $r$ := $\emptyset$ \\
\> \textbf{foreach} rule $x_i \rightarrow f(x_{i1}, \ldots, 
                x_{in})$ in $rules(i)$ \\
\> \> \textbf{if} exists rule $x_t \rightarrow f(x_{t1}, \ldots, 
                x_{tn})$ in $rules(t)$ \\
\> \> \> \textbf{for} $j$ = $1 .. n$ \\
\> \> \> \> $(r_j, x_j)$ := \func{rt}($x_{tj}$,$x_{ij}$, 
                  $P \cup \{ti(t,i)\}$) \\
\> \> \> \> \textbf{if} ($r_j = \ourtop$) \textbf{return} $(\ourtop, \_)$ \\
\> \> \> \textbf{endfor} \\
\> \> \> $r$ := $\{ti(t,i) \rightarrow f(x_1, \ldots, x_n)\} \cup r
 \cup r_1 \cup \cdots \cup r_n$ \\
\> \> \textbf{endif} \\
\> \textbf{endfor} \\
\> \textbf{return} $(r, ti(t,i))$ 
\end{tabbing}
\end{small}
\caption{Algorithm for computing the type instantiation 
grammar $\func{rt}(t,i)$\label{fig:rt}}

\end{figure}


\begin{example}\label{ex:rts}
Consider the types \texttt{list/1} and \texttt{habc} of
Example~\ref{ex:labcs} and instantiation \texttt{nelist/1} from
the program in Figure~\ref{fig:stack}.
Then ti-grammar \linebreak[2] 
$\func{rt}(\texttt{list(habc)},\texttt{nelist(old)})$ 
is 
\begin{grammar}
ti(\nt{list(habc)},\nt{nelist(old)}) & \rightarrow & [ ti(\nt{habc},\ntold) ~|~
ti(\nt{list(habc)},\nt{list(old)}) ] \\
ti(\nt{list(habc)},\nt{list(old)}) & \rightarrow & [] ~;~ [ ti(\nt{habc},\ntold) ~|~
ti(\nt{list(habc)},\nt{list(old)}) ] \\
ti(\nt{habc},\ntold) & \rightarrow & a ~;~ b ~;~ c ~;~ \#\texttt{var}\# 
\end{grammar}
while $\func{rt}(\texttt{list(T)},\texttt{nelist(ground)})$ is
\begin{grammar}
ti(\nt{list($T$)},\nt{nelist(ground)}) & \rightarrow & [ ti(T,\ntground) |
ti(\nt{list($T$)},\nt{list(ground)})] \\
ti(\nt{list($T$)},\nt{list(ground)}) & \rightarrow & [] ~;~ [ ti(T,\ntground) ~|~
ti(\nt{list($T$)},\nt{list(ground)}) ] \\
ti(T,\ntground) & \rightarrow & \$\ttground(T)\$
\end{grammar}
\end{example}

A ti-grammar is thus a regular tree grammar defined over the signature 
$$
\Sigma_{tree} \cup \{ \$\ttold(v)\$, \$\ttground(v)\$ 
~|~ v \in V_{type}\} \cup \{ \#\texttt{var}\#, \#\texttt{fresh}\# \}
$$
and non-terminal set
$$
\tau(\Sigma_{type} \cup \Sigma_{inst} \cup \{ti/2\}, V_{type})  
\cup \{ \nt{new} \}
$$

Note that by construction the partial ordering and meet and join
on tree grammars extend to ti-grammars including type parameters.
As mentioned before, type correctness guarantees that 
during mode checking we will only compare 
ti-grammars for the same type parameter $v \in V_{type}$. 
For this reason, we only need note
that $\func{rt}(v,\ttground) \preceq \func{rt}(v,\ttold)$
for a parameter $v \in V_{type}$, 
which follows from the construction since
$\oo \func{rt}(v,\ttground) \cc = \{ \$\texttt{ground}(v)\$ \}$
and 
$\oo \func{rt}(v,\ttold) \cc = \{ \$\texttt{ground}(v)\$, 
\$\texttt{old}(v)\$ \}$
and the meet and join operations follow in the natural way.

The operations that we perform on ti-grammars during mode checking
will be $\preceq$, abstract conjunction and abstract disjunction.
Abstract conjunction differs slightly from $\sqcap$ since we will
be changing variables with a \nt{new} ti-grammar to 
ti-grammars for bound values (whenever the variable becomes instantiated).
The abstract conjunction operation $\wedge$ is defined as:
$$
r_1 \wedge r_2 ~~ = ~~ \left\{ \begin{array}{l} 
                                 r_1, \mbox{~where~} r_2 = \nt{new} \\
                                 r_2, \mbox{~where~} r_1 = \nt{new} \\
                                 r_1 \sqcap r_2, \mbox{~otherwise}
                                 \end{array} \right.
$$

Abstract disjunction is again slightly different
from the $\sqcup$ operation. 
Since the compiler needs to know whether the value of
a variable 
is \texttt{new} or not, we need to ensure the abstract
disjunction operation does not create ti-grammars (other than $\ourtop$) 
in which this information is lost, i.e., grammars that include \#\texttt{fresh}\# as well as other terms.
The abstract disjunction operation $\vee$ is defined as:
$$
r_1 \vee r_2 ~~ = ~~ \left\{ \begin{array}{l} 
                                 r_1 \sqcup r_2, \mbox{~where~} r_1 \neq
                         \nt{new} \mbox{~and~}  r_2 \neq
                         \nt{new} \\
                                 \nt{new}, \mbox{~where~} r_1 = \nt{new} \mbox{~and~}   r_2 =
                         \nt{new}\\
                                 \ourtop, \mbox{~otherwise}
                                 \end{array} \right.
$$

Finally, we introduce the concept of
a \emph{type-instantiation state} (or \emph{ti-state})
$\{x_1 \mapsto r_1, \ldots, x_n \mapsto r_n\}$, which 
maps program variables to ti-grammars. Ti-grammars 
are used during mode checking
to store the possible
values of the program variables at each program point.
We can extend operations on ti-grammars to ti-states
over the same set of variables in the obvious pointwise manner.  
Given
ti-states $TI = \{x_1 \mapsto r_1, \ldots, x_n \mapsto
r_n\}$ and $TI' = \{x_1 \mapsto r'_1, \ldots, x_n \mapsto r'_n\}$ then:
\begin{itemize}
\item 
$TI
\preceq TI'$ iff $r_l \preceq r'_l$ for all $1 \leq l \leq n$, 
\item
$TI \wedge TI' =
\{ x_l \mapsto r_l \wedge r'_l ~|~ 1 \leq l \leq n \}$ and 
\item
$TI \vee TI' =
\{ x_l \mapsto r_l \vee r'_l ~|~ 1 \leq l \leq n \}$.
\end{itemize}

\section{Basic Mode Checking} \label{sec:basic}

Mode checking is a complex process
which aims to reorder body literals 
to satisfy the mode constraints provided by each mode
declaration. The aim of this is to be able to generate
specialized code for each mode declaration.
The code
corresponding to each mode declaration 
is referred to as a \emph{procedure}, and
calls to the original predicate are replaced by calls to the appropriate
procedure. 
Recall that before mode checking is applied the 
HAL compiler performs type checking (and inference) 
so that each program variable has a type, and the
program is guaranteed to be type correct.

\subsection{Well-moded programs}

We now define what it means for a HAL program to be well-moded.

The execution of a HAL program is performed on \emph{procedures}
which are predicates re-ordered for a particular mode.  
At run-time each type parameter has an associated 
ground type. 
For our purposes we assume a given \emph{type environment} $\theta$
(a ground type substitution) 
describes the run-time types associated with each type parameter.

A \emph{call} to a procedure $p/n$ in mode $p(c_1 \rightarrow s_1, \ldots, c_n \rightarrow s_n)$ is a type environment $\theta$ and a \emph{value}
$d_i$ for each argument $1 \leq i \leq n$. It follows from type correctness 
of the program that $d_i \in \oo \func{rt}(\theta(t_i),\ttold) \cc \cup 
\{\#\texttt{fresh}\#\}$ for each argument $1 \leq i \leq n$.

A program is \emph{input-output mode-correct} if any call 
to a predicate which is correct with respect to  the 
input instantiation for some mode 
declared for that predicate will only have answers that
are correct with respect to the output instantiation of that mode.
More formally, a program is 
\emph{input-output mode-correct} 
if for
each 
procedure $p/n$ with declared type $p(t_1, \ldots, t_n)$ in 
mode $p(c_1 \rightarrow s_1, \ldots, c_n \rightarrow s_n)$,  and
for any call of the form
$p(d_1, \ldots, d_n)$ with type environment $\theta$ 
such that $d_i \in \oo \func{rt}(\theta(t_i),c_i) \cc, 1 \leq i \leq n$,
it is the case that the resulting values $d'_1, \ldots, d'_n$ on 
success of the procedure are such that 
$d'_i \in \oo \func{rt}(\theta(t_i),s_i) \cc, 1 \leq i \leq n$.
In other words, the declared mode is satisfied by the code generated for the
procedure.

\begin{example}
For example the first
mode for predicate \texttt{pop/3}, defined in Example~\ref{fig:stack},
\begin{ttline}
:- mode pop(in,out,out) is semidet.
\end{ttline}
will be shown to be input-output mode-correct by showing that 
if the first argument to \texttt{pop/3} is ground at call time,
and the last two arguments \texttt{new},
then all three arguments will be ground on success of the
predicate.
\end{example}

A program is \emph{call mode-correct} if any call 
to a predicate which is correct with respect to the 
input instantiation for some mode 
declared for that predicate will only lead to calls to literals
within the definition of the predicate which are mode-correct.
More formally, a program is \emph{call mode-correct} if for
each procedure $p/n$ with declared type $p(t_1, \ldots, t_n)$ 
in mode  $p(c_1 \rightarrow s_1, \ldots, c_n \rightarrow s_n)$,   and
for any call of the form
$p(d_1, \ldots, d_n)$ with type environment $\theta$ 
such that $d_i \in \oo \func{rt}(\theta(t_i),c_i) \cc, 1 \leq i \leq n$,
it is the case that each call to a procedure 
$p'/n'$ with type (given by the occurrence in the definition of $p/n$) 
$p'(t'_1, \ldots, t'_{n'})$  in mode 
$p(c'_1 \rightarrow s'_1, \ldots, c'_{n'} \rightarrow s'_{n'})$ 
of the form
$p'(d'_1, \ldots, d'_n)$ is such that 
$d'_i \in \oo \func{rt}(\theta(t'_i),c'_i) \cc, 1 \leq i \leq n'$,
and any call to an equality of the form $x_1 = x_2$ is either a copy 
or unify, and any call to an equality of the form
$x = f(x_1, \ldots, x_n)$ is either a construct or deconstruct.
In other words each mode-correct call leads to only mode-correct calls.

\begin{example}\label{ex:dup}
Consider the following code which duplicates the top element of
the stack:
\begin{ttprog}
:- pred dupl(list(T), list(T)). \% duplicate top of stack \\
:- mode dupl(in(nelist(ground)), out(nelist(ground))) is det. \\
dupl(S0, S) :- S0 = [], S = []. \\
dupl(S0, S) :- push(S0, A, S), pop(S0, A, S1).
\end{ttprog}
Showing call mode-correctness for the procedure for 
\texttt{dupl/2} involves showing that any correct call to
\texttt{dupl/2} (that is with the first argument a non-empty ground list,
and its second argument \texttt{new})
will call \texttt{push/3} and \texttt{pop/3} with correct
input instantiations for one of their given modes,
and each equation must be either a construct or deconstruct.
\end{example}

A program is \emph{well-moded} if it is input-output mode-correct
and call mode-correct.

We shall now explain mode checking by showing how to 
check whether each program
construct is schedulable for a given ti-state $TI$ and, if so,
what the resulting ti-state $TI'$ is.
The scheduling also returns a goal 
illustrating the order of execution of conjunctions, and the mode
for each equation or predicate call. 
If the program construct is not schedulable for the 
given ti-state it may be
reconsidered after other constructs have been scheduled.
We assume that before
checking each construct for an initial ti-state $TI$, we
extend $TI$ so that any variable of type $t$ local to the construct is
assigned the ti-grammar \nt{new}.

\subsection{Equality}\label{sec:eq}

Consider the equality $x_1 = x_2$ where $x_1$ and $x_2$ are variables of
type $t$ and the current ti-state is $TI = \{x_1 \mapsto
r_1, x_2 \mapsto r_2\} \cup \emph{RTI}$ (where $\emph{RTI}$ is the ti-state for
the remaining variables).  The two standard modes of usage for such an
equality are \textbf{copy} (\texttt{:=}) and \textbf{unify} (\texttt{==}).
If exactly one of $r_1$ and $r_2$ is \nt{new} (say $r_1$), the 
\textbf{copy}
$x_1 := x_2$ can be performed and the resulting ti-state is $TI'
= \{x_1 \mapsto r_2, x_2 \mapsto r_2\} \cup \emph{RTI}$.  
If both are not \nt{new}
then \textbf{unify} $x_1 == x_2$ is performed and the resulting
instantiation is $TI' = \{x_1 \mapsto r_1 \wedge r_2, x_2 \mapsto r_1
\wedge r_2\} \cup \emph{RTI}$. 
If neither of the two modes of usage apply (i.e.\ 
both variables are new), the literal is not schedulable (although it might
become schedulable after automatic initialization, see Section \ref{sec:init}).

Consider the equality $x = f(x_1, \ldots, x_n)$ where $x, x_1,\ldots, x_n$
are variables with types $\{x \mapsto t, x_1 \mapsto t_1, \ldots, x_n
\mapsto t_n\}$ and current ti-state $TI = \{x \mapsto r, x_1
\mapsto r_1, \ldots, x_n \mapsto r_n\} \cup \emph{RTI}$. 
The two standard modes
of usage of such an equality are
\textbf{construct} (\texttt{:=}) and \textbf{deconstruct} (\texttt{=:}).
The \textbf{construct} mode applies if $r$ is \nt{new} and none of the
$r_j$ are \nt{new}. 
The resulting ti-state is $TI' = \{x
\mapsto r', x_1 \mapsto r_1, \ldots, x_n \mapsto r_n\} \cup \emph{RTI}$ 
where $r'$ is the ti-grammar 
defined by 
$\{a \rightarrow f(root(r_1), \ldots, root(r_n))\} \cup r_1 \cup 
\cdots  \cup r_n$, where $a$ is a new non-terminal,
(i.e.\ the grammar defining the terms constructible from 
an $f$ with arguments from $r_1,...,r_n$ respectively).
The \textbf{deconstruct} mode applies if each $r_j$ is \nt{new}
and $r$ is not \nt{new} and has no production rule
$root(r) \rightarrow \#\texttt{var}\# $
(which means it is definitely bound to some functor).
The resulting ti-state is 
$TI' = \{x \mapsto r, x_1 \mapsto r_1', \ldots ,
x_n \mapsto r_n'\} \cup \emph{RTI}$ where $r_1', \ldots, r_n'$
are defined below.
If $r$ has a production rule of the form 
$root(r) \rightarrow f(y_1, \ldots, y_n)$, then
the $r'_j = subg(y_j,r), 1 \leq i \leq n$.
If $r$ has no rule of this form, then 
the resulting ti-state is the same but with 
$r'_j = \bot, 1 \leq j \leq n$, 
indicating that the \textbf{deconstruct} must fail.
If some of the variables $x_j$ are
\nt{new} and some are not (say $x_{k_1},
\ldots, x_{k_m}$) the mode checking process decomposes the equality
constraint into a \textbf{deconstruct} followed by new equalities by
introducing fresh variables, e.g.\ $x = f(x_1, \ldots, \fresh_{k_j}, \ldots),
\ldots, x_{k_j} = \fresh_{k_j}, \ldots$.  These new equalities are
handled as above.  

Note that if $r = \nt{new}$ and some $r_i = \nt{new}$ 
then the literal is not schedulable
(although it might
become schedulable after automatic initialization, again 
see Section \ref{sec:init}).

\begin{example} 
Assume \texttt{X} and \texttt{Y} are ground lists, 
while \texttt{A} is \nt{new}.
Scheduling the goal \texttt{Y = [A|X]} 
results in the code \texttt{Y =:{~}[A|F], X == F}.
\end{example}

The above uses  of deconstruct are guaranteed to be safe at run-time 
and correspond to the modes of usage allowed by Mercury.
HAL, in addition to the above,
allows
the use of the \textbf{deconstruct} mode
when $x$ is \texttt{old} (i.e.\ $r$ contains a production rule
$root(r) \rightarrow \#\texttt{var}\#$).  
In this case we check whether 
$r$ has a production rule of the form 
$root(r) \rightarrow f(y_1, \ldots, y_n)$
and we proceed as in
the previous paragraph. Note that this is (the only place)
where the HAL mode system is not completely strong
(i.e.\ run-time mode errors can occur).
The following example illustrates the need for this behavior.

\begin{example}\label{ex:runtimeerror} 
Consider the types \texttt{abc/0} and \texttt{hlist/1} 
from Example~\ref{ex:labcs}, 
the following use of  \texttt{append/3} 
may not detect a mode error until run-time:
\begin{ttprog}
:- pred append(hlist(abc), hlist(abc), hlist(abc)). \\
:- mode append(oo, oo, no) is nondet. \\
append(X, Y, Z) :- X = [], Y = Z. \\
append(X, Y, Z) :- X = [A|X1], append(X1, Y, Z1), Z = [A|Z1].
\end{ttprog}
The equation \texttt{X = [A|X1]} 
is schedulable as a deconstruct since \texttt{X} is
\texttt{old}. However, if at run-time \texttt{X} is not bound 
when \texttt{append/3} is called, the
deconstruct will generate a run-time error since \texttt{A} is not a 
solver variable and, 
thus, it cannot be initialized. 
Note that if we did not allow
deconstruction on \texttt{old} variables then the above predicate would not 
pass mode checking thus preventing mode-correct goals like
\begin{ttline}
?- X = [a,b,c], init(Y), append(X,Y,Z).
\end{ttline}
from being compiled.
\end{example}

If we never allow Herbrand 
solver types to contain non-solver types (as in the example above), 
the problem cannot occur. 
This gap in mode checking seems unavoidable 
if we are to allow Herbrand solver types to contain non-solver types.
However, it seems that in practice this gap is not problematic: in most
programs, the possibility of a run-time mode error does not exist.
Whenever it does, the compiler emits a warning message. In fact, we
have never detected a run-time mode error.

\subsection{Predicates}

In this subsection 
we describe the scheduling of predicate calls so
that the resulting program after scheduling is call mode-correct.

Consider the predicate call $p(x_1, \ldots, x_n)$ where each
$x_i$ is a variable with type $t_i$.
Assume $p$ has the mode
declaration $p(c_1 \rightarrow s_1, \ldots, c_n \rightarrow s_n)$
where $c_j, s_j$
are the call and success instantiations, respectively, for argument $j$,
and the current ti-state is
$TI = \{x_1 \mapsto r_1, \ldots, x_n \mapsto r_n\} \cup \emph{RTI}$.

Note that the handling of polymorphic application is hidden here, since the
type $t_i$ of the variable $x_i$ is  type in the calling literal 
$p(x_1, \ldots, x_n)$, which may be more specific than the declared/inferred
type of argument $i$ of $p$. Because instantiations are separate from
types this is straightforwardly expressed by constructing the
ti-grammar
for the mode specific calling type $t_i$ and the appropriate instantiations.

The predicate call can be scheduled if for each $1 \leq j \leq n$ the
current ti-state restricts the $j$-th argument more than (defines a subset of) the calling
ti-state required for $p$, i.e.\ $r_j \preceq \func{rt}(t_j,c_j)$.  If the
predicate call 
is schedulable for this mode the new ti-state is $TI' = \{x_1
\mapsto r_1 \wedge \func{rt}(t_1,s_1), \ldots, x_n \mapsto r_n \wedge
\func{rt}(t_n,s_n)\} \cup \emph{RTI}$. 
The predicate call can also be
scheduled if for each $j$ such that $r_j \not\preceq \func{rt}(t_j,c_j)$ then 
$\func{rt}(t_j,c_j) = \nt{new}$. 
For each such $j$, the argument $x_j$ in
predicate call $p(x_1, \ldots,x_{j-1},x_j,x_{j+1},\ldots,x_n)$ is replaced
by $\fresh_j$, where $\fresh_j$ is a fresh new program variable, and the
equation $\fresh_j = x_j$ is added after the predicate call. 
Such ``extra'' modes are usually referred to as \emph{implied modes}.

\begin{example}
Consider the goal {\tt empty(S0)} for the program 
of Figure~\ref{fig:stack} where the type of \texttt{S0} is given by 
$\{ \texttt{S0} \mapsto \texttt{list(abc)}\}$ (which is more specific than
the declared type \texttt{list(T)})
and the current ti-state is $TI = \{
\texttt{S0} \mapsto \nt{new}\}$.
The two modes for \texttt{empty} (in expanded form) are 
\begin{ttline}
:- mode empty(ground -> ground) is semidet. \\
:- mode empty(new    -> ground) is det.
\end{ttline}
The first mode of \texttt{empty} cannot be scheduled since
$\nt{new} \not\preceq \func{rt}(\texttt{list(abc)},\texttt{ground})$,
but the second mode can be scheduled, since 
$\nt{new} \preceq \func{rt}(\texttt{list(abc)},\texttt{new}) = \nt{new}$.
\end{example}

If more than one mode of the same predicate is schedulable, in theory the compiler should
try each possibility. Unfortunately, this search may be too expensive. For
this reason, HAL (like Mercury) chooses one schedulable mode and commits to
it. This behavior might lead to the compiler failing to check a
mode-correct procedure (see Example~\ref{non-check}). In order to minimize
this risk, 
we choose a schedulable mode whose 
success ti-state $TI$ defined
as $\{x_1 \mapsto \func{rt}(t_1,s_1), \ldots, x_n \mapsto \func{rt}(t_n,s_n)\}$ is
minimal; that is, for each other schedulable mode with success 
ti-state $TI'$ it is the case that $TI' \not\preceq TI$. Note that 
there may be more than one mode with a minimal success ti-state. In 
the case that we have more than one mode with the same minimal success 
state then we use a mode with a
minimal call ti-state.

\begin{example}
Consider the scheduling of the goal
\texttt{pop(A,B,C)} where current ti-state is
$TI = \{ A \mapsto r_1, B \mapsto \nt{new}, C \mapsto r_3 \}$
where $r_1$ and $r_3$ are defined by the grammars
\begin{grammar}
r_1 & \rightarrow & [ r_2 ~|~ r_3 ] \\
r_2 & \rightarrow & b \\
r_3 & \rightarrow & []
\end{grammar}
That is $A = [b]$ and $C = []$.  Neither of the declared modes
for \texttt{pop}, shown below, are immediately applicable.
\begin{ttline}
:- mode pop(in,out,out) is semidet. \\
:- mode pop(in(nelist(ground)),out,out) is det. 
\end{ttline}
But both modes fit the conditions for an implied mode. 
Since the second mode has a more specific success ti-state
(the first argument is known to be non-empty)
it is chosen. 
The resulting code is \texttt{pop\_mode2(A,B,Fresh), Fresh = C},
where mode checking will then schedule the new equation appropriately.
\end{example}

The idea is to maintain as much instantiation information
as possible, thus restricting as little as possible the number of
schedulable modes for the remaining literals.
In our experience with compiling real programs 
this policy seems adequate to avoid any problems.
It is straightforward, but in practice too expensive,
to implement a complete search for all possible schedules.

\subsection{Conjunctions, Disjunctions and If-Then-Elses}

To determine if a conjunction $G_1, \ldots, G_n$ is schedulable for initial
ti-state $TI$ we choose the left-most goal $G_j$ which is schedulable for
$TI$ and compute the new ti-state $TI_j$. This default behavior schedules
goals as close to the programmer given left-to-right order as possible.  If the
state $TI_j$ assigns $\bot$ to any variable, then the subgoal $G_j$ must
fail and hence the whole conjunction is schedulable.  The resulting
ti-state $TI'$ maps all variables to $\bot$, and the final conjunction
contains all previously scheduled goals followed by \texttt{fail}.  
If $TI_j$ does
not assign $\bot$ to any variable we continue by scheduling the remaining
conjunction $G_1, \ldots, G_{j-1}, G_{j+1}, \ldots, G_n$ with initial
ti-state $TI_j$.  If all subgoals are eventually schedulable we have
determined both an order of evaluation for the conjunction and a final
ti-state. 

\begin{example}
Consider scheduling the goal
\begin{ttline}
Y = [U1|U2], U2 = [], X = [U1|U3].
\end{ttline}
where \texttt{X} is 
initially $\func{rt}(\texttt{list(T)},\texttt{ground})$, and the remaining
variables are \nt{new}.  The first literal is not schedulable and will
remain so until both \texttt{U1} and \texttt{U2} are no longer new. 
We consider then the
second literal, which is schedulable as a construct, thus changing the
type-instantiation of \texttt{U2} to
$\func{rt}(\texttt{list(T)},\texttt{elist})$.  Since the first
literal remains unschedulable, we consider the third literal which is
schedulable as a deconstruct, thus changing the type-instantiation of
\texttt{X},
\texttt{U1} and \texttt{U3} to
$\func{rt}(\texttt{list(T)},\texttt{nelist(ground)})$,
$\func{rt}(\texttt{T},\texttt{ground})$
and $\func{rt}(\texttt{list(T)},\texttt{ground})$, respectively. Since both
\texttt{U1} and \texttt{U2} 
are no longer new, the first literal is now schedulable as a
construct.  The resulting code is
\begin{ttline}
U2 := [], X =:{~}[U1|U3], Y := [U1|U2].
\end{ttline}
In the final ti-state the instantiation of $Y$ is given by the
tree-grammar
\begin{grammar}
\texttt{Y} & \rightarrow & [ ti(T,\nt{ground}) |
ti(\nt{list}(T),\nt{elist}) ] \\
ti(\nt{list}(T),\nt{elist}) & \rightarrow & [] \\ 
ti(T,\ntground) & \rightarrow & \$\ttground(T)\$
\end{grammar}
in other words it is a list of length exactly one.
\end{example}

To determine if a disjunction $G_1 ; \cdots ; G_n$ is schedulable for
initial ti-state $TI$ we check whether each subgoal $G_j$ is schedulable
for $TI$ and, if so, compute each resulting ti-state $TI_j$, obtaining the
final ti-state $TI' = \bigvee_{j \in \{1 .. n\}} TI_j$. 
If this
ti-state assigns $\ourtop$ to any variable or one of the disjuncts $G_j$
is not schedulable then the whole disjunction
is not schedulable.

To determine whether an if-then-else $G_i \rightarrow G_t ; G_e$ is schedulable
for initial ti-state $TI$, we determine first whether $G_i$
is schedulable for $TI$ with resulting ti-state $TI_i$.  If
not, the whole if-then-else is not schedulable. Otherwise, we try to
schedule $G_t$ in state $TI_i$ (resulting in state $TI_t$ say) and $G_e$ in
state $TI$ (resulting in state $TI_e$ say). 
The resulting ti-state is $TI' = TI_t \vee TI_e$. 
If one of $G_t$ or $G_e$ is not schedulable or $TI'$ includes $\ourtop$ the
whole if-then-else is not schedulable.
Note that the analysis of $G_i \rightarrow G_t ; G_e$ is identical to that of
$(G_i,G_t);G_e$ except that all goals of $G_i$ must be scheduled before those
of $G_t$.

\subsection{Mode Declarations}

In this subsection we discuss how mode-correctness is checked for
each mode declaration.

To check that a predicate with head $p(x_1, \ldots, x_n)$ and
 declared (or inferred) type
$\{x_1 \mapsto t_1, \ldots, x_n \mapsto t_n\}$
satisfies the mode declaration $p(c_1 \rightarrow s_1, \ldots, c_n
\rightarrow s_n)$, we build the initial ti-state
$TI = \{x_1 \mapsto \func{rt}(t_1,c_1), \ldots, x_n \mapsto \func{rt}(t_n,c_n)\}$. 
The body of the predicate 
is then analyzed starting from
the state $TI$.
The mode declaration is
correct if 
(a) everything is schedulable and
(b) if the final
ti-state is $TI' = \{x_1 \mapsto r_1', \ldots , x_n \mapsto r_n'\}$,
then  for each
argument variable $1 \leq i \leq n$, $r_i' \preceq \func{rt}(t_i,s_i)$.
If the body is not schedulable
or the resulting instantiations are not
strong enough, a mode error results. 
Note that (a) ensures that the predicate is call mode-correct
for that mode
while (b) ensures that it is input-output mode-correct.

\begin{example}
Consider mode checking of the following code from Example~\ref{ex:dup}
which makes use of the code in Figure~\ref{fig:stack}:
\begin{ttprog}
:- pred dupl(list(T), list(T)). \% duplicate top of stack \\
:- mode dupl(in(nelist(ground)), out(nelist(ground))) is det. \\
dupl(S0, S) :- S0 = [], S = []. \\
dupl(S0, S) :- push(S0, A, S), pop(S0, A, S1).
\end{ttprog}
We start by constructing 
the initial ti-state $TI = \{\texttt{S0} \mapsto gnelT,
\texttt{S} \mapsto \nt{new} \}$ where $gnelT = 
\func{rt}(\texttt{list(T)},\texttt{nelist(ground)})$ 
is the ti-grammar shown in Example~\ref{ex:rts}.  
Checking the first disjunct (rule) we have
\texttt{S0 = []} schedulable as a deconstruct.  
The resulting
ti-state assigns $\bot$ to \texttt{S0}, and thus 
the whole conjunction is
schedulable with  $TI_1 = \{\texttt{S0} \mapsto \bot, \texttt{S} \mapsto \bot \}$.
Checking the second disjunct, we first extend $TI$ to map 
\texttt{A} and \texttt{S1} to \nt{new}.
Examining the first literal \texttt{push(S0, A, S)}
we find that it is not schedulable since \texttt{A} has instantiation
\nt{new} and is required to be \ntground.
Examining the second literal \texttt{pop(S0, A,
S1)} we find that both 
modes declared for \texttt{pop/3} are
schedulable. Since the second mode has more specific success
instantiations, it is chosen and the ti-grammars for \texttt{A} and
\texttt{S1} become $\func{rt}(\texttt{T},\texttt{ground})$ and 
$\func{rt}(\texttt{list(T)}, \texttt{ground})$,
respectively.  
Now the first literal is schedulable obtaining for
\texttt{S} the
ti-grammar $gnelT$. 
Restricting to the original variables the
final ti-state is $TI_2 = \{\texttt{S0} \mapsto gnelT,
\texttt{S} \mapsto gnelT \}$.  
Taking the join $TI' = TI_1 \vee TI_2 = TI_2$.  
Checking this against the declared success
instantiations we find the declared mode is correct. The code generated for
the procedure is:
\begin{ttprog}
dupl\_mode1(S0, S) :- fail. \\
dupl\_mode1(S0, S) :- pop\_mode2(S0, A, S1), push\_mode1(S0, A, S).
\end{ttprog}
where \texttt{pop\_mode2/3} and \texttt{push\_mode1/3} are the procedures
associated with the second and first modes of the predicates, respectively.
\end{example}

Note that the HAL compiler's current mode analysis 
does not track variable dependencies
and thus it may obtain a final type-instantiation state weaker than
expected.  
\begin{example}\label{ex:lcint}
Consider the solver type \texttt{habc/0} of
Example~\ref{ex:labcs}.
The following program does not pass mode checking:
\begin{ttprog}
:- pred p(list(habc), habc). \\
:- mode p(list(old) -> ground, in) is semidet. \\
p(L, E) :- L = []. \\
p(L, E) :- L = [E1|L1], E = E1, p(L1, E).
\end{ttprog}
The first literal of the second rule 
is a deconstruct. After that deconstruct
variable \texttt{L} is never touched and hence its instantiation is
never updated; in particular it is not updated 
when the instantiation of 
\texttt{E1} and \texttt{L1} change. 
The inferred type-instantiation for \texttt{L} 
at the end of the second rule is thus
$\func{rt}(\texttt{list(habc)},\texttt{nelist(old)})$ rather than 
$\func{rt}(\texttt{list(habc)},\texttt{nelist(ground)})$. 
Hence, mode checking fails.
\end{example}

This could be overcome by adding a definite sharing
analysis and/or a dependency based groundness analysis to the 
mode checking phase. 
Whenever a variable which definitely
shares with another (through an equation $e$) is touched, we modify the
resulting ti-state as if the equation $e$ has been
rescheduled to update sharing variables.
This is (partially) 
implemented, for example, in the alias branch of the Mercury
compiler.

\section{Automatic Initialization}\label{sec:init}

As mentioned before, constraint solvers must provide an initialization
procedure (\texttt{init/1}) for their solver type. This procedure takes a
solver variable with instantiation \texttt{new} and returns it with
instantiation \texttt{old}, after initializing whatever data-structures (if
any) the solver needs. 

Many of the predicates exported by constraint solvers
(including most constraints) require the solver variables appearing as
arguments to be already initialized.
Thus, explicit initializations for local variables may
need to be introduced. Not only is this a tedious exercise for the programmer,
it may even be impossible for multi-moded predicate definitions
since each mode may require different initialization instructions.  
Therefore, the HAL mode checker
automatically inserts variable initializations. In particular,
whenever a literal cannot be scheduled because there is a
requirement for an argument of type $t$ 
to be $\func{rt}(t,\texttt{old})$ when it is \nt{new} 
and $t$ is a solver type, then the \texttt{init/1} predicate for
type $t$ can be inserted to make the literal schedulable.

\begin{example}
Assume we have an integer solver with solver type  \texttt{cint/0}.
\begin{ttprog}
:- pred length(list(cint), int). \\
:- mode length(out(list(old)), in) is nondet. \\
:- mode length(in(list(old)), out) is det. \\
length(L, N) :- L = [], N = 0. \\
length(L, N) :- L = [X|L1], +(N1,1,N), N > 0, length(L1, N1).
\end{ttprog}
where the predicate \texttt{+(X,Y,Z)} models $X + Y = Z$ and 
requires at least two arguments to be ground on call and all arguments are
ground on return.

For the first mode \texttt{L = [X|L1]} cannot be scheduled as a construct
until \texttt{X} has a ti-grammar 
different from \nt{new}. Hence, \texttt{X}
needs to be initialized.  
In the second mode \texttt{L = [X|L1]} can be scheduled as
a deconstruct and thus no initialization is needed. 
The two resulting procedures are:

\begin{ttprog}
length\=\_mode1(L, N) :-  (L := [], N == 0 \\
                  \> ;  +$_{outinin}$(N1, 1, N), N > 0, length\_mode1(L1, N1), init(X), L := [X|L1]). \\
length\_mode2(L, N) :-  (L == [], N := 0 \\
                  \> ;   L =:{~}[X|L1], length\_mode2(L1, N1),
                  +$_{ininout}$(N1, 1, N), N > 0). 
\end{ttprog}

\noindent
where we have rewritten the call to \texttt{+/3} to show the mode more
clearly (\texttt{+$_{outinin}$} indicates that the first argument is 
\texttt{out} and the
rest are \texttt{in}, \texttt{+$_{ininout}$} indicates that the 
third argument is \texttt{out}
and the other arguments are \texttt{in}).  
\end{example}

Unfortunately, unnecessary initialization may slow down execution and introduce
unnecessary variables (when it interacts with implied modes).  Hence, we
would like to only add those initializations required so that mode checking
will succeed.  The HAL mode checker implements this by first trying to mode
the procedure without allowing initialization. If this fails it will start
from the previous partial schedule looking for the leftmost unscheduled
literal $l$ which can be scheduled by initializing variables which (a) have
either a solver type or a parameter type (e.g.~$v \in V_{type}$) 
and (b) do not appear in an unscheduled
literal to the left 
which equates them to a term (if so, chances are the equation will
become a construct and no initialization is needed). If such an $l$ is
found the appropriate initialization calls are inserted before $l$, and 
then scheduling continues once more trying to schedule
without initialization.  
If no $l$ is found the whole conjunct is not schedulable.  This two phase
approach is applied at each conjunct level individually.

\begin{example}
Consider the following program where \texttt{cint/0} is a solver type:
\begin{ttprog} 
:- instdef evenlist(T) -> ([] ; [T|oddlist(T)]). \\
:- instdef oddlist(T) -> [T|evenlist(T)]. \\
\\
:- pred pairlist(list(cint),int). \\
:- mode pairlist(out(evenlist(old)),in) is nondet. \\
pairlist(L,N) :- N = 0, L = []. \\
pairlist(L,N) :- N > 0, +(N1,1,N), L = [V|L1], L1 = [V|L2], pairlist(L2,N1).
\end{ttprog}
In the first phase all literals in the second
rule are schedulable except \texttt{L = [V|L1]} and \texttt{L1 =
[V|L2]} which can be neither a 
construct (\texttt{V}, \texttt{L1} and \texttt{L2} are new) nor a 
deconstruct (both \texttt{L} and \texttt{L1} are new). 
In the second phase we examine the two remaining unscheduled
literals: the second literal can be scheduled by initializing \texttt{V}. Once this is done the
first literal can be scheduled obtaining:
\begin{ttprog}
pairlist(L,N) :- N == 0, L := []. \\
pairlist(L,N) :- \= N > 0, +$_{outinin}$(N1,1,N), pairlist(L2,N1), \\
\> init(V), L1 := [V|L2], L := [V|L1].
\end{ttprog}
\end{example}

Many other different initialization heuristics could be applied. We are
currently investigating more informed policies which give the right tradeoff
between adding constraints as early as possible, and delaying constraints
until they can become tests or assignments.

\section{Higher-Order Objects}\label{sec:higher}

Higher-order programming is particularly important
in HAL because it is the mechanism
used to implement dynamic scheduling, which is vital in CLP
languages for extending and combining constraint solvers.  
Higher-order programming 
introduces two new kinds of literals: construction
of higher-order objects and higher-order calls.  

A \emph{higher-order object} is constructed
using an equation of the form 
$h = p(x_{1}, \ldots, x_k)$ where
$h, x_{1}, \ldots, x_k$ are variables and 
$p$ is an $n$-ary predicate with
$n \geq k$. 
The variable $h$ is referred to as a \emph{higher-order object}.
\emph{Higher-order calls} are literals of the form \texttt{call}$(h, x_{k+1},
\ldots, x_n)$ where $h, x_{k+1}, \ldots, x_n$ are variables.  Essentially,
the \texttt{call} literal supplies the $n-k$ arguments missing from the
higher-order object $h$.

In order to represent types and instantiations for
higher-order objects we need to extend the languages of
type and instantiation expressions.
The \emph{higher-order type} of a higher-order object $h$
constructed in the previous paragraph 
is of the form $pred(t_{k+1}, \ldots, t_n)$ 
where $pred/(n-k)$ is a new special type constructor and
$t_{k+1}, \ldots, t_n$ are types. 
It provides the types of the $n-k$ arguments missing from $h$. 
The \emph{higher-order instantiation} of $h$ 
is of the form $pred(c_{k+1} \rightarrow s_{k+1}, \ldots, c_n
\rightarrow s_n)$\footnote{In reality, the determinism information 
also appears in the
higher-order instantiation; for simplicity we ignore it here.} 
where $pred/(n-k)$ is a new
special instantiation construct and $c_j \rightarrow s_j$ give the call and
success instantiations of argument $j$ respectively. 
It provides the modes of the $n-k$ arguments missing from $h$.
Note that for the first time we allow \ttnew{} instantiations appearing
inside instantiation expressions (since they will often be call
instantiations). But their appearance is restricted to the outermost arguments 
of higher-order instantiations.

Now we must extend the 
$\func{rt}(t,i)$ operation to handle higher-order types and instantiations.
Let us first consider the case in which 
$i$ is the higher-order instantiation 
$pred(c_{k+1} \rightarrow s_{k+1}, \ldots, c_n \rightarrow s_n)$. 
If $t$ is the higher-order type $pred(t_{k+1}, \ldots,
t_n)$ then $\func{rt}(t,i)$ returns the grammar
\begin{grammar}
ti(t,i) & \rightarrow & \ipred(root(tc_{k+1}), root(ts_{k+1}),
\ldots, root(tc_n), root(ts_n))
\end{grammar}
together with the grammars 
$tc_{k+1}, \ldots, tc_n, ts_{k+1}, \ldots, ts_n$
where $tc_j = \func{rt}(t_j,c_j)$ and $ts_j = \func{rt}(t_j,s_j)$.
If $t$ is not a higher-order type or has the wrong arity 
then $\func{rt}(t,i) = \ourtop$, indicating an error.
The new constant \ipred{} simply collects the
call and success ti-grammars for the higher-order object's 
missing arguments.

The extension of $\func{rt}(t,i)$ for the case of base
instantiations $i$ is similar to the treatment of type parameters.  
A higher-order object can be \texttt{new} or 
\texttt{ground}, but if it is \texttt{old} this is identical
to \texttt{ground} since higher-order objects never have an
attached solver.
$\func{rt}(pred(t_1,\ldots,t_n), \texttt{new})$ is treated as
before (i.e.~it creates a \nt{new} ti-grammar).
Similarly $\func{rt}(pred(t_1,\ldots,t_n), \texttt{ground})$ 
generates a production rule using a new constant 
\gpred{} of the form
$$
ti(pred(t_1,\ldots,t_n), \texttt{ground}) \rightarrow
\gpred
$$
$\func{rt}(pred(t_1,\ldots,t_n), \texttt{old})$ generates the same
grammar (since it is equivalent).
Since we will only compare the higher-order ti-grammar
against other ti-grammars for the same type we can safely omit
the information about the argument types ($t_1, \ldots, t_n$).

The new constant \gpred{} 
acts like $\$\texttt{ground}(v)\$$ 
but it can also be compared with more complicated ti-grammars 
(with production rules for function symbol \ipred)
of the same type.  The full code for $\func{rt}(t,i)$ is given in the appendix.

\begin{example}\label{ex:map}
Consider the following code:
\begin{ttprog}
:- pred map(pred(T1,T2), list(T1), list(T2)). \\
:- mode map(in(pred(in,out) is det), in, out) is det. \\
map(H, [], []). \\
map(H, [A|As], [B|Bs]) :- call(H,A,B), map(H,As,Bs). \\
\\
:- typedef sign -> (neg ; zero ; pos). \\
\\
:- pred mult(sign, sign, sign). \\
:- mode mult(in, in, out) is det. \\
\\
?- H1 = mult(pos), map(H1, [neg,zero,pos], L1).
\end{ttprog}
The \texttt{map/3} predicate takes a higher-order predicate with two
missing arguments of parametric types \texttt{T1} and \texttt{T2} 
and modes \texttt{in} and  \texttt{out},
respectively. 
The ti-grammar describing the input instantiation of
the first argument of \texttt{map/3} is the grammar with 
root $a_1 = ti(\nt{pred(T1,T2)}, \nt{pred(in,out)})$, defined by
\begin{grammar}
a_1 & \rightarrow & 
\ipred(ti(\nt{T1},\ntground),ti(\nt{T1},\ntground),\ntnew,ti(\nt{T2},\ntground)) \\
ti(\nt{T1},\ntground) & \rightarrow & \$\ttground(T1)\$ \\
ti(\nt{T2},\ntground) & \rightarrow & \$\ttground(T2)\$ \\
\ntnew & \rightarrow & \#\texttt{fresh}\#
\end{grammar}

This predicate is applied to a list of \texttt{T1}s, returning a list
of \texttt{T2}s. 
The literal \texttt{H1 = mult(pos)} 
constructs a higher-order object which 
multiplies the sign of its first argument 
by \texttt{pos}, returning the result in its
second argument.
The type-instantiation of \texttt{H1},
$\func{rt}(\texttt{pred(sign,sign)}, \texttt{pred(in,out)})$,
is the grammar with root $a_2 = ti(\nt{pred(sign,sign)},
\nt{pred(in,out)})$ and rules:
\begin{grammar}
a_2 & \rightarrow &
\ipred(ti(\nt{sign},\ntground),ti(\nt{sign},\ntground),\nt{new}, 
ti(\nt{sign},\ntground)) \\
ti(\nt{sign},\ntground) & \rightarrow & neg ~;~ zero ~;~ pos \\
\ntnew & \rightarrow & \#\texttt{fresh}\#
\end{grammar}
\end{example}

We need to extend the ordering $\preceq$ to 
higher-order type-instantiations
as well as the operations $\wedge$ and $\vee$.  
Two higher-order ti-grammars $r$ and $r'$ defined with rules
\begin{grammar}
root(r) & \rightarrow & \ipred(xc_1, xs_1, \ldots, xc_n, xs_n)
\end{grammar}
and
\begin{grammar}
root(r') & \rightarrow & \ipred(xc'_1, xs'_1, \ldots, xc'_n, xs'_n)
\end{grammar}
satisfy $r \preceq r'$ iff for $i=1,\ldots,n$, 
$subg(xc'_i, r') \preceq subg(xc_i, r)$ and 
$subg(xs_i, r) \preceq subg(xs'_i, r')$.
Intuitively, if $r \preceq r'$, then any higher-order $\texttt{call}(r',\ldots)$ should 
be replaceable by 
$\texttt{call}(r,\ldots)$. 
For this to work, two conditions must be fulfilled. First,
$r$ must be able to deal with any
values that $r'$ can deal with (and perhaps more). Thus, 
$subg(xc'_i, r') \preceq subg(xc_i, r)$. 
And second, $r$ must return the same values
as $r'$ or less. Thus, $subg(xs_i, r) \preceq subg(xs'_i, r')$. 
For more details see 
the example below.

We define $r \preceq \func{rt}(\texttt{pred}(t_1,\ldots,t_n),\ttground)$ 
for any ti-grammar $r$ of the appropriate type except \ntnew.
The full definition of $\preceq$ is given in the appendix.
The $\wedge$ and $\vee$ operations follow naturally from the
ordering, and are given in the appendix.

\begin{example}
Consider the following code and goal:
\begin{ttprog}
:- typedef abc -> a ; b ; c. \\
:- instdef ab -> a ; b. \\
\\
:- pred ho1(abc,abc). \\
:- mode ho1(in(ab),out(ab)) is det.\\
ho1(A,B) :- A = B.\\
\\
:- pred ho2(abc,abc). \\
:- mode ho2(in,out) is det.\\
ho2(A,B) :- A = a, B = b.\\
ho2(A,B) :- A = b, B = c.\\
ho2(A,B) :- A = c, B = a.\\\\
?- HO1 = ho1, HO2 = ho2, (HO = HO1 ; HO = HO2).
\end{ttprog}
During scheduling of the disjunction, 
the ti-grammar for
\texttt{HO1} is \func{rt}(\texttt{pred(abc,abc)},
\texttt{pred(in(ab),out(ab))}), i.e.:
\begin{grammar}
\nt{ho1}& \rightarrow &  \ipred(\nt{gndab},\nt{gndab},\nt{new},\nt{gndab}) \\
\nt{gndab} & \rightarrow & a ~;~ b \\
\nt{new}   & \rightarrow & \#\texttt{fresh}\#
\end{grammar}
and the ti-grammar for \texttt{HO2} is
\begin{grammar}
\nt{ho2}& \rightarrow &  \ipred(\nt{gndabc},\nt{gndabc},\nt{new},\nt{gndabc}) \\
\nt{gndab} & \rightarrow & a ~;~ b ~;~ c\\
\nt{new}   & \rightarrow & \#\texttt{fresh}\#
\end{grammar}
The abstract disjunction of these two grammars to build the
ti-grammar for \texttt{HO}  gives
\begin{grammar}
\nt{ho} & \rightarrow &  \ipred(\nt{gndab},\nt{gndabc},\nt{new},\nt{gndabc}) \\
\end{grammar}
Notice the call ti-grammars have been abstractly conjoined.  This
illustrates the contravariant nature of calling instantiations
of higher-order predicates.
The higher-order object in \texttt{HO} can only be safely applied to
an input \texttt{a} or \texttt{b} since it may be predicate \texttt{ho1}.
It can only be guaranteed to give output \texttt{a},\texttt{b} or \texttt{c}
since it may be predicate \texttt{ho2}.
\end{example}

\subsection{Scheduling Higher-Order}\label{sec:sched-ho}

Intuitively, a higher-order equation $h = p(x_{1}, \ldots, x_k)$ is
schedulable if $h$ is \nt{new} and $x_{1}, \ldots, x_k$ are at least as
instantiated as the call instantiations of one of the modes declared
for $p/n$. If this is true for more than one mode, we again choose one schedulable 
mode (using the same criteria used for calls to first order predicates) 
and commit to it. 
If it is not true for any mode, the equation is delayed until the
arguments become more instantiated.  Formally,
let the current ti-state be 
$TI = \{h \mapsto r, x_1
\mapsto r_1, \ldots, x_k \mapsto r_k\} \cup \emph{RTI}$
and the types $\{x_{1} \mapsto t_1, \ldots, x_k \mapsto t_k\}$.
Let the (declared or inferred) predicate 
type of $p/n$ be $p(dt_1, \ldots, dt_n)$, then (because of type correctness)
we have that there exists $\theta$ such that $\theta(dt_j) = t_j$.

Consider the declared mode 
$p(c_1 \rightarrow s_1, \ldots, c_n \rightarrow s_n)$. 
The higher-order equation is schedulable if $r = \nt{new}$ 
and for each $1 \leq j \leq k,~ r_j \preceq
\func{rt}(t_j,c_j) ~\wedge~ r_j \neq \nt{new}$.  
The resulting ti-state is
$$
\{h \mapsto r', 
x_1 \mapsto r_1, \ldots, x_k \mapsto r_k\} \cup \emph{RTI}.
$$
where $tc_j = \func{rt}(\theta(dt_{j}),c_{j})$ and 
$ts_j = \func{rt}(\theta(dt_{j}),s_{j})$
for $k+1 \leq j \leq n$ and 
$r' = \{ a \rightarrow 
\ipred(root(tc_{k+1}), root(ts_{k+1}), \ldots, root(tc_{n}),
root(ts_{n}))\}$, where $a$ is a new non-terminal,
together with the grammars for 
$tc_{k+1}, ts_{k+1}, \ldots, tc_{n}, ts_{n}$.

Note that the instantiation of each $x_j$ is
unchanged and, in fact, will not be updated even when $h$ is
called. This is because in general we cannot ensure when or if the
call has actually been made. 
As a result, mode checking with higher-order objects can be imprecise.
In particular, if one of the
$r_j$ is \nt{new} we may not know if it becomes initialised or not 
since we do not know if the call to $h$ which will initialise it has been made. Since 
we must be able to precisely track when a variable has become 
initialised, we do not allow a call to be scheduled if this
is the case (hence the $r_j \neq \nt{new}$ condition above).  

A higher-order call \texttt{call}$(h, x_{k+1}, \ldots, x_n)$ is schedulable
if $x_{k+1}, \ldots, x_n$ are at least as instantiated as the call
instantiations of the arguments
of the higher-order type-instantiation previously assigned to $h$. If this is
not true, the call is delayed until the arguments become more instantiated.
Formally, let the current ti-state be 
$TI = \{h \mapsto r,
x_{k+1} \mapsto r_{k+1}, \ldots, x_n \mapsto r_n\} \cup \emph{RTI}$. 
The call is
schedulable if $r$ has a production rule of the form 
$$
root(r) \rightarrow 
\ipred(xc_{k+1}, xs_{k+1},  \ldots, xc_n, xs_n)
$$
and for each $j \in k+1, ..., n, ~ r_j \preceq subg(xc_j, r)$.  
The resulting instantiation is 
$TI' = \{h \mapsto r, x_{k+1} \mapsto
r_{k+1} \wedge subg(xs_{k+1},r), \ldots, x_n \mapsto r_n \wedge subg(xs_n,r)\} 
\cup \emph{RTI}$.
Just as for normal predicate calls, implied modes are also possible
where if, for example, $xc_l$ is \nt{new}, 
we can replace $x_l$ with a fresh
variable $\fresh_l$ and a following equation $\fresh_l = x_l$.
And, if necessary, the mode checker will add calls to initialise
solver variables.

\begin{example}
Consider the following code and assume all goals are schedulable in the
order written:
\begin{ttprog}
:- instdef only\_a -> a. \\
:- modedef abc2a -> (ground -> only\_a). \\
\\
:- pred p(abc, abc, abc) \\
:- mode p(abc2a, in, out(only\_a)) is semidet. \\
\\
?- $G_1$, p(A,B,C), $G_2$. \\
?- $G_1$, H = p(A), call(H,B,C), $G_2$. 
\end{ttprog}
The two queries would appear to have identical 
effects. However, mode checking for the second
goal will not determine that the 
instantiation for \texttt{A} becomes \texttt{only\_a}
by the time it reaches goal $G_2$.
Assuming \texttt{A} was \texttt{ground} before  \texttt{H = p(A)},
then the type-instantiation of \texttt{H} is the grammar with
root $x = ti(pred(\nt{abc},\nt{abc}),pred(\nt{in},\nt{out(only\_a)}))$ and
rules:
\begin{grammar}
x 
& \rightarrow &
\ipred(ti(\nt{abc},\ntground),ti(\nt{abc},\ntground),
\nt{new}, ti(\nt{abc}, \nt{only\_a})) \\
ti(\nt{abc},\ntground) & \rightarrow & a ~;~ b ~;~ c \\
ti(\nt{abc},\nt{only\_a}) & \rightarrow & a \\
\nt{new} & \rightarrow & \#\texttt{fresh}\# 
\end{grammar}
Of course in this case it is obvious that the predicate is being called
before $G_2$, and so it could be inferred that the instantiation of 
\texttt{A}
was \nt{only\_a} at that point.
However, in the usual case such analysis is harder, since the construction
of a higher-order term and its eventual execution are usually performed
in different predicates.
Indeed, in general it is impossible to know at compile time whether at
a given program point the higher-order predicate has been executed or not.
\end{example}

\section{Polymorphism and Modes}\label{sec:poly}

Polymorphic predicates are very useful because they can be
used for different types. Unfortunately, mode information can be lost since
only the base instantiations \texttt{ground}, \texttt{new}, and
\texttt{old} can be associated with type parameters. 

\begin{example}
Consider the interface to the stack data type defined in Figure
\ref{fig:stack} and the following program:
\begin{ttprog}
:- pred q(abc). \\
:- mode q(in) is semidet. \\
:- mode q(in(only\_a)) is det. \\
\\
?- empty(S0), I0 = a, push(S0, I0, S1), pop(S1, I, S2), q(I). 
\end{ttprog}
Although list \texttt{S1} is indeed a list only containing items \texttt{a}
this information is lost after executing \texttt{push} since the output
instantiation declared for this predicate is simply ground. Because of
this, the first mode of predicate \texttt{q/1} will be selected for
literal \texttt{q(I)}, thus losing the information 
that \texttt{q(I)} could not fail.
\end{example}

This loss of instantiation information for arguments to polymorphic 
predicates may have severe consequences for higher-order objects
because the base instantiation \texttt{ground} applied to polymorphic
code does not contain enough information for the higher-order object to be
used (called). 
\begin{example}
Consider the following goal using code from Figure~\ref{fig:stack}
and Example~\ref{ex:map}:
\begin{ttprog}
?- empty(S0), I0 = mult(pos), push(S0,I0,S1), pop(S1,I,S2), map(I,[neg],S). 
\end{ttprog}
When item \texttt{I} is extracted from the list its ti-grammar is
$\func{rt}(t,\nt{ground})$ where $t$ is type $pred(\nt{sign},\nt{sign})$.
As a result, it cannot be used in
\texttt{map} since its mode and determinism
information has been lost, i.e.~the check
$\func{rt}(t,\nt{ground}) \preceq \func{rt}(t,\nt{pred(in,out)})$ fails.
\end{example}

We could overcome the above problem by having a special version of each
stack predicate to handle the higher-order predicate case. But this requires
modifying the \texttt{stack} module, defeating the idea of an abstract data
type. Also, this modification is required for each mode of the higher-order
object that the programmer wishes to make use of.  
Clearly, this is not an attractive proposition.

Our approach is to use polymorphic type information to recover the lost mode 
information.  This is an example of ``Theorems for
Free''~\cite{wadler}: since the polymorphic code can only
equate terms with polymorphic
type, it cannot create instantiations and, thus, the output instantiations of
polymorphic arguments must result from the calling instantiations of
non-output arguments. Hence, they have to be at least as instantiated as the
join of the input instantiations.

\subsection{Polymorphic Mode Checking}

To recover instantiation information we extend mode checking for procedures
with polymorphic types to take into account the extra mode information that
is implied by the polymorphic type.  
Consider the predicate call $p(x_1,
\ldots, x_n)$ where $x_1, \ldots, x_n$ are variables with type $\{x_1
\mapsto t_1, \ldots, x_n \mapsto t_n\}$ and 
current ti-state 
$TI = \{x_1 \mapsto r_1, \ldots, x_n \mapsto r_n\} \cup \emph{RTI}$.  
Suppose the predicate 
type declared (or inferred) for $p$ is $p(dt_1, \ldots, dt_n)$.  Note that
because of type correctness there exists the
type substitution $\theta$ where $\theta(dt_j) = t_j$.

\begin{figure}[t]

\begin{small}
\begin{tabbing} 
xx \= xx \= xx \= xx \= xx \= \kill
\func{collect\_set}($r_1$,$r_2$,$P$) \\
\> $x_1$ := $root(r_1)$; $x_2$ := $root(r_2)$ \\
\> \textbf{if} ($(x_1,x_2) \in P$) \textbf{return} $\emptyset$\\
\> \textbf{if} ($x_1 = \nt{new}$)  \textbf{return} $\emptyset$ \\
\> \textbf{if} ($x_1 \rightarrow \svars{v} \in r_1$) 
        \textbf{return} $\{(\ttold,v,r_2)\}$ \\
\> \textbf{if} ($x_1 \rightarrow \sgrounds{v} \in r_1$)
        \textbf{return} $\{(\ttground,v,r_2)\}$ \\
\> $M$ := $\emptyset$ \\
\> \textbf{foreach} rule $x_1 \rightarrow f(x_{11}, \ldots, x_{1n})$ in $r_1$ \\
\> \> \textbf{if} exists rule $x_2 \rightarrow f(x_{21}, \ldots, x_{2n})$ in
$r_2$ \\
\> \> \> \textbf{for} $i$ := $1 .. n$ \\
\> \> \> \> $M$ := $M~ \cup$ \func{collect\_set}($subg(x_{1i},r_1)$, $subg(x_{2i},r_2)$, $P \cup \{(x_1,x_2)\}$) \\
\> \textbf{return} $M$ \\
\end{tabbing}
\end{small}
\caption{Algorithm for collecting the type-instantiations that match
 type parameters.\label{fig:collect}}
\label{collect_set}

\end{figure}

Assume the literal is schedulable for mode declaration 
$p(c_1 \rightarrow s_1, \ldots, c_n \rightarrow s_n)$.  
We proceed by matching the ti-trees $\func{rt}(dt_j, c_j)$ 
against the current instantiations $r_j$ in a
process analogous to the matching that occurs in the meet function.
Note that $\func{rt}(dt_j, c_j)$ is the ti-grammar which contains information
on the positions of type parameters in the declared type of $p$.

Consider the function \func{collect\_set}($r_1,r_2,\emptyset$),
defined in Figure~\ref{collect_set}, which returns the set of triples
$(\ttold,v,r')$ and $(\ttground,v,r')$ obtained by collecting
each ti-grammar, $r'$, in $r_2$ 
matching occurrences of $\svars{v}$ and  $\sgrounds{v}$ in 
$r_1$. 
Let 
$M = \cup_{j=1}^n$ \func{collect\_set}($\func{rt}(dt_j,c_j),r_j,\emptyset$).  
We will use this 
information to compute the success instantiations as follows: since the
only success type-instantiation information for elements of parametric type
$v$ must come from its call type-instantiations, we can safely assume that
any success type-instantiation is at least as instantiated as the join
(upper bounds) of the calls.

Note that when determining ground success information, we need only
consider ground calling instantiations, since ground success instantiations
cannot result from old call instantiations.
On the other hand, for old success information, we need to consider both old
and ground calling instantiations, since old success instantiations can
result from either.
Hence the following definitions for $ground(v,M)$ and $old(v,M)$, which
compute upper bounds on success instantiations for $v$ based on the call
instantiation information collected in $M$:
\begin{eqnarray*}
ground(v,M) & = & \bigvee \{ r ~|~ (\ttground,v,r) \in M \} \\
old(v,M) & = & \bigvee \{ r ~|~ (\ttground,v,r) \in M \mbox{~or~}
(\ttold,v,r) \in M\}
\end{eqnarray*}
Because the literal is schedulable for the given mode we know
that no $r_i$ contains \nt{new} for any $t$. 
Thus,  the abstract disjunctions in  
$ground(v,M)$ and $old(v,M)$ never lead to $\ourtop$.

Let $ps_j$ be the result of replacing in 
$\func{rt}(dt_j, s_j)$ 
each non-terminal $x$ with productions of the form
$$
x \rightarrow \sgrounds{v} ~;~ \svars{v}
$$
by $root(old(v,M))$ and removing the rules for
$x$, and replacing each non-terminal $x$ with productions of the form
$$
x \rightarrow \sgrounds{v}
$$
by $root(ground(v,M))$ and removing the rules for $x$, and finally adding
the rules in $old(v,M)$ and $ground(v,M)$.
The new ti-state
resulting after scheduling the polymorphic literal is $TI' = \{x_1 \mapsto
r_1 \wedge ps_1, \ldots, x_n \mapsto r_n \wedge ps_n\} \cup \emph{RTI}$.

\begin{example} \label{ex:hopush}
Assume we are scheduling the \texttt{push/3} literal in the
goal using code from Figure~\ref{fig:stack} and Example~\ref{ex:map}:
\begin{ttline}
?- empty(S0), I0 = mult(pos), push(S0,I0,S1), pop(S1,I,S2), map(I,[neg],S). 
\end{ttline}
for current ti-state 
$\{\texttt{S0} \mapsto r_3, \texttt{I0} \mapsto r_4\}$,
the remaining variables being \nt{new},
where $r_3$ is the grammar
\begin{grammar}
ti(\nt{list(sign)},\nt{elist}) & \rightarrow & [] \\
\end{grammar}
and $r_4$ is the grammar with root 
$a = ti(pred(\nt{sign},\nt{sign}),pred(\nt{in},\nt{out}))$ defined by
\begin{grammar}
a & \rightarrow &
        \ipred(ti(\nt{sign},\ntground),ti(\nt{sign},\ntground),
                          \ntnew,ti(\nt{sign},\ntground)) \\
ti(\nt{sign},\ntground) & \rightarrow & neg ~;~ zero ~;~ pos
\end{grammar}
The ti-grammars 
defined by the declared type and mode declarations 
for the first two arguments of
\texttt{push/3} are: $r_5 = \func{rt}(\texttt{list(T)},\texttt{ground})$ 
or the grammar
\begin{grammar}
ti(\nt{list(T)},\nt{ground}) & \rightarrow & [] ~;~ [ ti(\nt{T},\ntground)
~|~ti(\nt{list(T)},\nt{ground}) ] \\
ti(\nt{T},\ntground) & \rightarrow & \$\texttt{ground}(T)\$ 
\end{grammar}
and, $r_6 = \func{rt}(\nt{T},\nt{ground})$, the grammar
\begin{grammar}
ti(\nt{T},\ntground) & \rightarrow & \$\texttt{ground}(T)\$ 
\end{grammar}
The literal is schedulable and the matching process determines that
$\func{collect\_set}(r_5,r_3) = \emptyset$, 
$\func{collect\_set}(r_6,r_4) = \{(\nt{ground},\nt{T}, r_4)\}$
and $M = \{(\nt{ground},\nt{T}, r_4)\}$. 
The improved analysis 
determines that extra success instantiation 
($ps_3$) 
for the third argument (\texttt{S1}) by improving
$\func{rt}(\texttt{list(T)},\texttt{nelist(ground)})$ which is
\begin{grammar}
ti(\nt{list(T)},\nt{nelist(ground)} & \rightarrow & [ ti(\nt{T},\ntground)
~|~ ti(\nt{list(T)},\nt{list(ground)} ] \\
ti(\nt{list(T)},\nt{list(ground)} & \rightarrow & [] ~;~ [ ti(\nt{T},\ntground)
~|~ ti(\nt{list(T)},\nt{list(ground)} ] \\
ti(\nt{T},\ntground) & \rightarrow & \$\texttt{ground}(T)\$ 
\end{grammar}
replacing the last rule by $r_4$ and occurrences of
$ti(\nt{list(T)},\nt{list(ground)}$ by $root(r_4) = a$ obtaining
\begin{grammar}
ti(\nt{list(T)},\nt{nelist(ground)} & \rightarrow & [ {a}
~|~ ti(\nt{list(T)},\nt{list(ground)} ] \\
ti(\nt{list(T)},\nt{list(ground)} & \rightarrow & [] ~;~ [ a
 ~|~ ti(\nt{list(T)},\nt{list(ground)} ] \\
a & \rightarrow &
        \ipred(ti(\nt{sign},\ntground),ti(\nt{sign},\ntground),
                          \ntnew,ti(\nt{sign},\ntground)) \\
ti(\nt{sign},\ntground) & \rightarrow & neg ~;~ zero ~;~ pos
\end{grammar}
Note that the mode information of the higher-order term has been preserved.
The mode checking for the call to \texttt{pop/3} 
will similarly preserve the higher-order mode information, and
the original goal will be schedulable.
\end{example}

The interaction between polymorphic mode analysis and higher-order
constructs and calls is in fact slightly more complicated than discussed
previously. This is because higher-order objects allow us to give arguments
to a predicate in a piecewise manner. This affects the execution of
\func{collect\_set} which was collecting the set $M$ over all predicate
arguments simultaneously. In order to handle these accurately we need to
store the information from $M$ found during the higher-order object
construction, to be used in the higher-order call. That is, we need to store
$ground(v,M)$ and $old(v,M)$ for each type parameter $v$ appearing in the
remaining arguments as part of the ti-grammar for the higher-order object.

An alternative approach used by the HAL compiler is to update the success
instantiations stored in the ti-grammar of the higher-order object based on
the extra information from polymorphism.  When the call to the higher-order
polymorphic predicate is analyzed, the matching process also matches the
success instantiations of the higher-order object to recover the previous
matching information.

\section{Conclusions and Future Work}\label{sec:concl}

The ultimate aim of mode checking is to ensure that the compiler has correct
instantiation information at every program point in order to
allow program optimization.  It is reasonably 
straightforward (but laborious) 
to show that the mode checking defined in this paper
ensures that the resulting program has input-output 
and call correctness.
Some subtle points that arise are as follows. 
First, 
it is an invariant that any ti-grammar (or sub-grammar) $r$ occurring in the
mode checking process that contains rule
$root(r) \rightarrow \#\texttt{var}\#$ must be equivalent 
to $\func{rt}(t,\ttold)$ for some $t$, 
which means that when variables are 
bound indirectly (through shared variables) the 
correctness of the ti-state is maintained. 
Second, if a procedure is input-output correct for
its declared type, then it is also input-output correct for any
instance of the type. This follows from the limited possibilities
for manipulating objects of variable type (essentially copying
and testing equality).

This means that compiler optimizations can safely be applied.
The only mode error that may be detected at run-time
arises from situations explained in Section~\ref{sec:eq} and
Example~\ref{ex:runtimeerror}.\footnote{Note this does not invalidate the
input-output or call correctness for the remainder of the program.}
The compiler emits warnings when such a possibility exists.

We have described for the first time 
mode checking for CLP languages, such as HAL,
which have strong typing and re-orderable clause bodies, 
and described the algorithms currently used in the HAL compiler.
The actual implementation of these algorithms in the HAL 
compiler is considerably more
sophisticated than the simple presentation here.  Partial schedules are
computed and stored and accessed only when enough new instantiation
information has been created to reassess them. Operations such as $\preceq$ 
are tabled and hence many operations are simply a 
lookup in a table. 
We have found mode checking is efficient enough for a practical compiler.
For the compiler compiling itself 
(29000 lines of HAL code in 27 highly interdependent 
modules compiled in 15 mins 20 secs)
mode checking requires 
16.4\% of overall compile time.
While compiling the libraries
(4600 lines of HAL code in 12 almost independent modules compiled in 47 secs)
it takes 13.1\% of overall compile time.
And compiling a suite of small to medium size benchmarks 
(6200 lines of HAL code in 67 modules compiled in 183 secs)
it takes 13.0\% of overall compile time,

There is considerable scope for future work. 
One aim is to strengthen mode checking. 
We plan to add tracking of aliasing and groundness dependencies. 
Another problem is that currently
HAL (like Mercury) never undoes a feasible choice of ordering the literals.
This can lead to correctly moded programs not being checkable as in
Example \ref{non-check}.  In practice this behavior is rare, but we
would like to explore more complete strategies.

\begin{example}
\label{non-check}
Consider the following declarations and goal:
\begin{ttprog}
:- pred p(list(int),list(int)). \\
:- mode p(out,out) is det. \\
:- mode p(in(evenlist(ground)),out(evenlist(ground))) is det. \\
\\
:- pred q(list(int)). \\
:- mode q(out(evenlist(ground))) is det. \\
\\
:- pred r(list(int)). \\
:- mode r(in(evenlist(ground))) is det. \\
\\
?- p(L0, L1), q(L0), r(L1).
\end{ttprog}
The first two literals of the goal are schedulable in the 
order given, as 
\texttt{p\_mode1(L0, L1), q\_mode1(L2), L2 = L0}
but then \texttt{r(L1)} is not schedulable (the list \texttt{L1}
may not be of even length). There is a feasible schedule:
\texttt{q\_mode1(L0), p\_mode2(L0, L1), r\_mode1(L1)}
which is missed by both HAL and Mercury, since they don't undo the
feasible schedule for the first two literals.
In order to avoid this problem HAL allows the user to name modes
of a predicate and hence specify exactly which mode is required.
\end{example}

A second aim is to improve the efficiency of the reordered code,
by, for instance, reducing the number of initializations.
The final aim is to provide mode inference as well as mode 
checking---the ability to reorder body literals  makes this a potentially very expensive
process.

\subsection*{Acknowledgements}

We would like to thank Fergus Henderson and Zoltan Somogyi for discussions
of the Mercury mode system, and their help in modifying Mercury to support
HAL features. We would also like to thank
the anonymous referees whose suggestions have enormously improved the paper.


\small

\appendix
\section{Algorithms}

In this appendix we give full versions of
the tree operations mentioned in the paper.  The basic tree operations are
relatively straightforward, but new kinds of nodes for solver variables,
polymorphic types and higher-order terms complicate this somewhat.
Recall that we assume we are dealing with type correct programs, hence the
operations make use of this to avoid many redundant comparisons. For example
when comparing the order of two ti-grammars, then if one is a predicate type,
the other must be an identical predicate type.

The ordering relation $r_1 \preceq r_2$ on two ti-grammars 
is defined as the result of $\func{lt}(r_1, r_2, \emptyset)$. 

\begin{small}
\begin{tabbing} 
xx \= xx \= xx \= xx \= xx \= xx \= xxxxxxxxxxxxxxxxxxxxx \= xxxxxxx \= \kill
\func{lt}($p_1$, $p_2$, $P$) \\
\> \textbf{if} ($p_2 = \ourtop$) \textbf{return true} \\
\> \textbf{if} ($p_1 = \ourtop$) \textbf{return false} \\
\> \textbf{if} ($(root(p_1),root(p_2)) \in P$) \textbf{return true} \\
\> \textbf{if} ($p_2 = \nt{new} \mbox{ and } p_1 \neq \nt{new}$) \textbf{return false} \\
\> \textbf{case}: \\
\> \> $p_1 = \ntnew$: \textbf{return} $(p_2 = \nt{new})$ \\
\> \> $root(p_1) \rightarrow \svars{v} \in p_1$: 
        \> \> \> \> \> \%\% $p_1 =  \func{base}(v,\ttold)$ \\
\> \> \> \textbf{return} $root(p_2) \rightarrow \svars{v} \in p_2$ \\
\> \> $root(p_1) \rightarrow \sgrounds{v} \in p_1$: 
        \> \> \> \> \> \%\% $p_1 = \func{base}(v,\ttground)$ \\
\> \> \> \textbf{return} $root(p_2) \rightarrow \sgrounds{v} \in p_2$ \\
\> \> $root(p_1) \rightarrow \gpred \in p_1$: 
        \> \> \> \> \> \%\% $p_1 = \func{base}(pred(t_1,\ldots,t_n),\ttground)$ \\
\> \> \>  \textbf{return} $root(p_2) \rightarrow \gpred \in p_2$ \\
\> \> $root(p_1) \rightarrow \ipred(tc_1, ts_1, \ldots, tc_n, ts_n)
         \in p_1$:  \> \> \> \> \> \> \%\% non-base higher-order ti \\
\> \> \> \textbf{if} ($root(p_2) \rightarrow \gpred \in p_2$) 
        \textbf{return} \textbf{true} \\
\> \> \> \textbf{let} $root(p_2) \rightarrow 
        \ipred(tc'_1, ts'_1, \ldots, tc'_n, ts'_n) \in p_2$ \\
\> \> \> \textbf{for} $i$ := $1..n$ \\
\> \> \> \> \textbf{if} ($\neg
\func{lt}(subg(tc'_i,p_2),subg(tc_i,p_1),P \cup \{(root(p_1),root(p_2))\})$)
\textbf{return} \textbf{false} \\
\> \> \> \> \textbf{if} ($\neg
\func{lt}(subg(ts_i,p_1),subg(ts'_i,p_2),P \cup \{(root(p_1),root(p_2))\})$)
\textbf{return} \textbf{false} \\
\> \> \> \textbf{endfor} \\
\> \> \> \textbf{return true} \\
\> \> \textbf{default}: \\
\> \> \> \textbf{foreach} $root(p_1) \rightarrow f(x_1, \ldots, x_n)
\in p_1$ \\
\> \> \> \> \textbf{if} ($\exists root(p_2) \rightarrow f(x'_1, \ldots,
x'_n) \in p_2$) \\
\> \> \> \> \> \textbf{for} $i$ := $1 ..n$ \\
\> \> \> \> \> \> \textbf{if} ($\neg \func{lt}(subg(x_i, p_1), subg(x'_i,
p_2), P \cup \{(root(p_1),root(p_2))\})$) \textbf{return false} \\
\> \> \> \> \> \textbf{endfor} \\
\> \> \> \> \textbf{else} \textbf{return} \textbf{false} \\
\> \> \> \textbf{endfor} \\
\> \> \> \textbf{return true} 
\end{tabbing}
\end{small}

The abstract conjunction operation
$r_1 \wedge r_2$ on two ti-grammars
is defined as the first element of the pair returned by
$\func{conj}(r_1,r_2, \emptyset)$.

\begin{small}
\begin{tabbing} 
xx \= xx \= xx \= xx \= xx \= xx \= xxxxxxxxxxxxxxxxxxxxx \= xxxxxxx \= \kill
\func{conj}($p_1$,$p_2$,$P$) \\
\> \textbf{if} ($p_1 = \ourtop$) \textbf{return} $(\ourtop,\_)$\\ 
\> \textbf{if} ($p_2 = \ourtop$) \textbf{return} $(\ourtop,\_)$\\ 
\> \textbf{if} ($p_2 = \nt{new}$) \textbf{return} $(p_1,root(p_1))$\\ 
\> \textbf{if} ($meet(root(p_1),root(p_2)) \in P$) 
\textbf{return} $(\emptyset,meet(root(p_1),root(p_2)))$\\
\> \textbf{case}: \\
\> \> $p_1 = \nt{new}$: \textbf{return} ($p_2$, $root(p_2)$)  \\
\> \> $root(p_1) \rightarrow \svars{v} \in p_1$: 
        \> \> \> \> \> \%\% $p_1 =  \func{base}(v,\ttold)$ \\
\> \> \>  \textbf{return} ($p_2$,$root(p_2)$) \\
\> \> $root(p_1) \rightarrow \sgrounds{v} \in p_1$: 
        \> \> \> \> \> \%\% $p_1 = \func{base}(v,\ttground)$ \\
\> \> \>  \textbf{return} ($p_1$,$root(p_1)$) \\
\> \> $root(p_1) \rightarrow \gpred \in p_1$: 
     \> \> \> \> \> \%\% $p_1 = \func{base}(pred(t_1,\ldots,t_n),\ttground)$ \\
\> \> \>  \textbf{return} ($p_2$,$root(p_2)$) \\
\> \> $root(p_1) \rightarrow \ipred(tc_1, ts_1, \ldots, tc_n, ts_n)
         \in p_1$:  \> \> \> \> \> \> \%\% non-base higher-order ti \\
\> \> \> \textbf{if} ($root(p_2) \rightarrow \gpred \in p_2$) \textbf{return} ($p_1$,$root(p_1)$) \\
\> \> \> \textbf{let} $root(p_2) \rightarrow 
        \ipred(xc'_1, xs'_1, \ldots, xc'_n, xs'_n) \in p_2$ \\
\> \> \> \textbf{for} $i$ := $1..n$ \\
\> \> \> \> $(tc_i, xc''_i)$ := $\func{disj}(subg(xc_i,p_1),subg(xc'_i,
p_2),P)$ \\
\> \> \> \> $(ts_i, xs''_i)$ := $\func{conj}(subg(xs_i,p_1),subg(xs'_i,
p_2),P)$ \\
\> \> \> \> \textbf{if} ($tc_i = \ourtop \mbox{ or } ts_i = \ourtop$)
                \textbf{return} $(\ourtop,\_)$ \\
\> \> \> \textbf{endfor} \\
\> \> \> $p$ := \= $\{ meet(root(p_1), root(p_2))
\rightarrow \ipred(xc''_1, xs''_1, \ldots, xc''_n, xs''_n)\} \cup$ \\
\> \> \> \> $tc_1  \cup \cdots \cup tc_n \cup ts_1 \cup \cdots \cup ts_n$ \\
\> \> \> \textbf{return} $(p,  meet(root(p_1),root(p_2)))$ \\
\> \> \textbf{default}: \\
\> \> \> $p$ := $\emptyset$ \\
\> \> \> \textbf{foreach} $root(p_1) \rightarrow f(x_1, \ldots, x_n)
\in p_1$ \\
\> \> \> \> \textbf{if} ($\exists root(p_2) \rightarrow f(x'_1, \ldots,
x'_n) \in p_2$) \\
\> \> \> \> \> \textbf{for} $i$ := $1 .. n$ \\
\> \> \> \> \> \> $(p''_i, x''_i)$ := $\func{conj}(subg(x_i,p_1),
subg(x'_i, p_2),  P \cup \{meet(root(p_1),root(p_2))\})$) \\
\> \> \> \> \> \> \textbf{if} ($p''_i = \ourtop$)
                \textbf{return} $(\ourtop,\_)$  \\
\> \> \> \> \> \textbf{endfor} \\
\> \> \> \> \> $p$ := $meet(root(p_1),root(p_2)) \rightarrow f(x''_1,
\ldots, x''_n) \cup p \cup p''_1 \cup \cdots \cup p''_n$ \\
\> \> \> \textbf{endfor} \\
\> \> \> \textbf{return} $(p, meet(root(p_1),root(p_2)))$  
\end{tabbing}
\end{small}

The abstract disjunction operation
$r_1 \vee r_2$ on two ti-grammars, 
is defined as the first element of the pair returned by
$\func{disj}(r_1,r_2, \emptyset)$.

\begin{small}
\begin{tabbing} 
xx \= xx \= xx \= xx \= xx \= xx \= xxxxxxxxxxxxxxxxxxxxx \= xxxxxxx \= \kill
\func{disj}($p_1$,$p_2$,$P$) \\
\> \textbf{if} ($p_1 = \ourtop$) \textbf{return} $(\ourtop,\_)$\\ 
\> \textbf{if} ($p_2 = \ourtop$) \textbf{return} $(\ourtop,\_)$\\ 
\> \textbf{if} ($p_1 = \nt{new} \mbox{ and } p_2 = \nt{new}$) 
        \textbf{return} $(\{\nt{new} \rightarrow
\#\texttt{fresh}\#\},\nt{new})$\\ 
\> \textbf{if} ($p_2 = \nt{new}$) \textbf{return} $(\ourtop,\_)$\\ 
\> \textbf{if} ($\exists join(root(p_1),root(p_2)) \in P$) 
\textbf{return} $(\emptyset,join(root(p_1),root(p_2)))$\\
\> \textbf{case}: \\
\> \> $p_1 = \nt{new}$: \textbf{return} ($\ourtop$, $\_$)  \\
\> \> $root(p_1) \rightarrow \svars{v} \in p_1$: 
        \> \> \> \> \> \%\% $p_1 =  \func{base}(v,\ttold)$ \\
\> \> \>   \textbf{return} ($p_1$,$root(p_1)$) \\
\> \> $root(p_1) \rightarrow \sgrounds{v} \in p_1$: 
        \> \> \> \> \> \%\% $p_1 = \func{base}(v,\ttground)$ \\
\> \> \> \textbf{return} ($p_2$,$root(p_2)$) \\
\> \> $root(p_1) \rightarrow \gpred \in p_1$:  
     \> \> \> \> \> \%\% $p_1 = \func{base}(pred(t_1,\ldots,t_n),\ttground)$ \\
\> \> \>   \textbf{return} ($p_1$,$root(p_1)$) \\
\> \> $root(p_1) \rightarrow \ipred(xc_1, xs_1, \ldots, xc_n, xs_n)
         \in p_1$:  \> \> \> \> \> \> \%\% non-base higher-order ti \\
\> \> \> \textbf{if} ($root(p_2) \rightarrow \gpred \in p_2$) 
        \textbf{return} ($p_2$,$root(p_2)$) \\
\> \> \> \textbf{let} $root(p_2) \rightarrow 
        \ipred(xc'_1, xs'_1, \ldots, xc'_n, xs'_n) \in p_2$ \\
\> \> \> \textbf{for} $i$ := $1..n$ \\
\> \> \> \> $(tc_i, xc''_i)$ := $\func{conj}(subg(xc_i,p_1),subg(xc'_i,
p_2),P)$ \\
\> \> \> \> $(ts_i, xs''_i)$ := $\func{disj}(subg(xs_i,p_1),subg(xs'_i,
p_2),P)$ \\
\> \> \> \> \textbf{if} ($tc_i = \ourtop \mbox{ or } ts_i = \ourtop$)
                \textbf{return} $(\ourtop,\_)$ \\
\> \> \> \textbf{endfor} \\
\> \> \> $p$ := \= $\{join(root(p_1), root(p_2))
\rightarrow \ipred(xc''_1, xs''_1, \ldots, xc''_n, xs''_n)\} \cup$ \\
\> \> \> \> $tc_1 \cup \cdots \cup tc_n \cup ts_1 \cup \cdots \cup ts_n$ \\
\> \> \> \textbf{return} $(p,  join(root(p_1),root(p_2)))$ \\
\> \> \textbf{default}: \\
\> \> \> $p$ := $\emptyset$ \\
\> \> \> \textbf{foreach} $root(p_1) \rightarrow f(x_1, \ldots, x_n)
\in p_1$ \\
\> \> \> \> \textbf{if} ($\exists root(p_2) \rightarrow f(x'_1, \ldots,
x'_n) \in p_2$) \\
\> \> \> \> \> \textbf{for} $i$ := $1 .. n$ \\
\> \> \> \> \> \> $(p''_i, x''_i)$ := $\func{disj}(subg(x_i,p_1),
subg(x'_i, p_2),  P \cup \{join(root(p_1),root(p_2))\})$) \\
\> \> \> \> \> \> \textbf{if} ($p''_i = \ourtop$)
                \textbf{return} $(\ourtop,\_)$  \\
\> \> \> \> \> \textbf{endfor} \\
\> \> \> \> \> $p$ := $\{join(root(p_1),root(p_2)) \rightarrow f(x''_1,
\ldots, x''_n)\} \cup p \cup p''_1 \cup \cdots \cup p''_n$ \\
\> \> \> \> \textbf{else} \\
\> \> \> \> \> $p$ := \= $\{join(root(p_1),root(p_2) \rightarrow f(x_1, \ldots,
x_n)\} \cup p \cup$  \\
\> \> \> \> \> \>  $subg(x_1, p_1) \cup \cdots \cup subg(x_n,p_1)$ \\
\> \> \> \> \textbf{endif} \\
\> \> \> \textbf{endfor} \\
\> \> \> \textbf{foreach}  $root(p_2) \rightarrow f(x'_1, \ldots, x'_n)
\in p_2$ \\
\> \> \> \> \textbf{if} ($\neg \exists root(p_1) \rightarrow f(x_1, \ldots,
x_n) \in p_1$) \\
\> \> \> \> \> $p$ := $\{join(root(p_1),root(p_2) \rightarrow f(x'_1, \ldots,
x'_n)\} \cup p \cup$  \\
\> \> \> \> \> \>  $subg(x'_1, p_2) \cup \cdots \cup subg(x'_n,p_2)$ \\
\> \> \> \textbf{endfor} \\
\> \> \> \textbf{return} $(p, join(root(p_1),root(p_2)))$  
\end{tabbing}
\end{small}

The \func{rt} operation constructs a ti-grammar
from a type $t$ and instantiation $i$ and 
is defined as the first element in the pair resulting from
$\func{rt}(t, i, \emptyset)$. 

\begin{small}
\begin{tabbing}
xx \= xx \= xx \= xx \= xx \= xx \= xx \= xx \= \kill
\func{rt}($t$,$i$,$P$) \\
\> \textbf{if} ($\exists ti(t,i) \in P$) 
                \textbf{return} $(\emptyset,ti(t,i)))$\\
\> \textbf{case}: \\
\> \> $i$ is a base instantiation: \textbf{return}  
        \func{base}($t$,$i$,$P$) \\
\> \> $i = pred(c_1 \rightarrow s_1, \ldots, c_n \rightarrow s_n)$: \\
\> \> \> \textbf{if} ($t \neq pred(t_1, \ldots, t_n)$) \textbf{return}
$(\ourtop, \_)$ \\
\> \> \> \textbf{let} $t$ be of the form $pred(t_1, \ldots, t_n)$ \\
\> \> \> \textbf{for} $j$ = $1 .. n$ \\
\> \> \> \> $(tc_j, xc_j)$ := \func{rt}($t_j$, $c_j$, $P$) \\ 
\> \> \> \> $(ts_j, xs_j)$ := \func{rt}($t_j$, $s_j$, $P$) \\ 
\> \> \> \> \textbf{if} ($tc_j = \ourtop \mbox{ or } ts_j = \ourtop$)
                \textbf{return} $(\ourtop, \_)$ \\
\> \> \> \textbf{endfor} \\
\> \> \> $r$ := $\{ti(t,i) \rightarrow \ipred(xc_1,xs_1,
\ldots, xc_n, xs_n)\} \cup$  \\
\> \> \> \> \> $tc_1 \cup \cdots \cup tc_n \cup ts_1 \cup \cdots \cup ts_n$ \\
\> \> \> \textbf{return} $(r, ti(t,i))$ \\ 
\> \> \textbf{default}: \\
\> \> \> \textbf{if} ($t \in V_{type}$) \textbf{return} $(\ourtop, \_)$ \\
\> \> \> $r$ := $\emptyset$ \\
\> \> \> \textbf{foreach} $x \rightarrow f(x_{i1}, \ldots, 
                x_{in}) \in rules(i)$ \\
\> \> \> \> \textbf{if} ($\exists x' \rightarrow f(x_{t1}, \ldots, 
                x_{tn}) \in rules(t)$) \\
\> \> \> \> \> \textbf{for} $j$ = $1 .. n$ \\
\> \> \> \> \> \> $(r_j, x_j)$ := \func{rt}($x_{tj}$, $x_{ij}$, 
                $P \cup \{ti(x_{tj},x_{ij})\}$) \\
\> \> \> \> \> \> \textbf{if} ($r_j = \ourtop$) \textbf{return} $(\ourtop,\_)$ \\
\> \> \> \> \> \textbf{endfor} \\
\> \> \> \> \> $r$ := $\{ti(t,i) \rightarrow f(x_1, \ldots, x_n)\} \cup r
 \cup r_1 \cup \cdots \cup r_n$ \\
\> \> \> \> \textbf{endif} \\
\> \> \> \textbf{endfor} \\
\> \> \> \textbf{return} $(r, ti(t,i))$ 
\end{tabbing}

\begin{tabbing}
xx \= xx \= xx \= xx \= xx \= xx \= xx \= xx \= \kill
\func{base}($t$,$base$,$P$) \\
\> \textbf{if} ($base = \ttnew$) \textbf{return} ($\{\nt{new} \rightarrow
\#\texttt{fresh}\#\}, \nt{new}$) \\
\> \textbf{if} ($ti(t,base) \in P$) \textbf{return} ($\emptyset$,$ti(t,base)$) \\
\> \textbf{if} ($t \in V_{type}$): \\
\> \> \textbf{if} ($base = \ttground$) 
        \textbf{return} 
        ($\{ti(t,\ntground) \rightarrow \sgrounds{t}\}, ti(t,\ntground)$) \\
\> \> \textbf{else} \textbf{return} ($\{ti(v,\ntold) \rightarrow
\sgrounds{v} ~;~ \svars{v}\}, 
                ti(v, \ntold)$) \\ 
\> \textbf{else if} ($t$ is of the form $pred(t_1, \ldots, t_n)$) \\
\> \>  \textbf{return} (\{$ti(pred(t_1, \ldots, t_n),\ttground) \rightarrow \gpred$\}, 
           $ti(pred(t_1, \ldots, t_n),\ttground)$ ) \\
\> \textbf{else} \\
\> \> $r$ := $\emptyset$ \\
\> \> \textbf{foreach} $x \rightarrow f(t_{1}, \ldots, 
                t_{n})$ in $rules(t)$ \\
\> \> \> \textbf{for} $j \in 1 .. n$ \\
\> \> \> \> $(r_j,x_j)$ := \func{base}($t_j$,$base$,$P \cup \{
ti(t,base) \}$) \\
\> \> \> \textbf{endfor} \\
\> \> \> $r$ := $\{ti(t,base) \rightarrow f(x_1, \ldots, x_n)\} \cup r
 \cup r_1 \cup \cdots \cup r_n$ \\
\> \> \textbf{endfor} \\
\> \> \textbf{if} ($base = \ttold$ and $t$ is a solver type) \textbf{then} $r$ := $\{ti(t,base)
\rightarrow \#\texttt{var}\#\} \cup r$ \\
\> \> \textbf{return} $(r,ti(t,base))$

\end{tabbing}
\end{small}

\end{document}